\documentclass[reprint,superscriptaddress,amsmath,amssymb,aps,prd,nofootinbib]{revtex4-2}
\makeatletter
\def\@bibdataout@aps{%
 \immediate\write\@bibdataout{%
  @CONTROL{%
   apsrev42Control%
   \longbibliography@sw{%
    ,author="48",editor="1",pages="0",title="0",year="1"%
   }{%
    ,author="48",editor="1",pages="0",title="",year="1"%
   }%
  }%
 }%
 \if@filesw
  \immediate\write\@auxout{\string\citation{apsrev42Control}}%
 \fi
}%
\makeatother

\usepackage{graphicx}
\graphicspath{{figures/}}
\usepackage{bm}
\usepackage{placeins}
\usepackage{xcolor}
\definecolor{darkblue}{rgb}{0,0,0}
\definecolor{darkgreen}{rgb}{0,0,0}
\definecolor{newgreen}{rgb}{0,0,0}
\makeatletter
\@ifl@t@r\fmtversion{2023-06-01}{%
 \@ifl@t@r\fmtversion{2026-06-01}{%
  \def\label#1{\@bsphack
   \begingroup
   \UseHookWithArguments{label}{1}{#1}%
   \protected@write\@auxout{\let\label\@gobble@om\let\index\@gobble@som\let\glossary\@gobble@om}%
     {\string\newlabel{#1}{{\@currentlabel}{\thepage}%
     {\@currentlabelname}{\@currentHref}{\@kernel@reserved@label@data}}}%
   \endgroup
   \@esphack
  }%
 }{%
  \def\label#1{\@bsphack
   \begingroup
   \UseHookWithArguments{label}{1}{#1}%
   \protected@write\@auxout{}%
     {\string\newlabel{#1}{{\@currentlabel}{\thepage}%
     {\@currentlabelname}{\@currentHref}{\@kernel@reserved@label@data}}}%
   \endgroup
   \@esphack}%
 }%
 \let\ltx@label\label
}{}
\makeatother
\usepackage{hyperref}

\newcommand{\kv}{\mathbf{k}}
\newcommand{\qv}{\mathbf{q}}
\newcommand{\xv}{\mathbf{x}}
\newcommand{\rv}{\mathbf{r}}
\newcommand{\nv}{\hat{\mathbf{n}}}
\newcommand{\vv}{\hat{\mathbf{v}}}
\newcommand{\Bt}{\tilde{B}}
\newcommand{\zt}{\tilde{\zeta}}
\newcommand{\Plin}{P_{\rm lin}}
\newcommand{\Pnw}{P_{\rm nw}}
\newcommand{\Mpc}{\,h^{-1}\,{\rm Mpc}}
\newcommand{\hMpc}{\,h\,{\rm Mpc}^{-1}}
\newcommand{\Hf}{\mathcal{H}f}
\newcommand{\gaunt}[6]{\mathcal{G}_{#1\,#2\,#3}^{#4\,#5\,#6}}
\newcommand{\Klam}{\mathcal{K}_\Lambda}
\newcommand{\bG}{b_{\mathcal{G}_2}}

\newcommand{\rev}[1]{#1}
\colorlet{BLUE}{blue}

\begin{document}

\title{Anisotropic three-point statistics of the peculiar velocity and galaxy density fields: modelling in Fourier and configuration space}

\author{M. Guidi}
\email{massimo.guidi6@unibo.it}
\affiliation{Dipartimento di Fisica e Astronomia ``Augusto Righi'', Alma Mater Studiorum Universit\`a di Bologna, via Piero Gobetti 93/2, I-40129 Bologna, Italy}
\affiliation{INAF -- Osservatorio di Astrofisica e Scienza dello Spazio di Bologna, via Piero Gobetti 93/3, I-40129 Bologna, Italy}
\author{A. Farina}
\affiliation{INAF -- Osservatorio Astronomico di Brera, via Brera 28, I-20122 Milano, Italy}
\affiliation{INFN -- Sezione di Genova, via Dodecaneso 33, I-16146 Genova, Italy}
\author{A. Nusser}
\affiliation{Department of Physics, Technion -- Israel Institute of Technology, Haifa 3200003, Israel}
\author{E. Branchini}
\affiliation{Dipartimento di Fisica, Universit\`a di Genova, via Dodecaneso 33, I-16146 Genova, Italy}
\affiliation{INFN -- Sezione di Genova, via Dodecaneso 33, I-16146 Genova, Italy}
\affiliation{INAF -- Osservatorio Astronomico di Brera, via Brera 28, I-20122 Milano, Italy}
\date{\today}

\begin{abstract}
\textcolor{darkblue}{By the equivalence principle, large-scale peculiar velocities provide a kinematical probe of the matter distribution independent of the biased galaxy distribution. Their three-point correlation with galaxies assesses the nonlinear coupling between cosmic flows and the surrounding density field. We develop a tree-level model of the anisotropic correlation between one line-of-sight peculiar velocity and two galaxy densities in redshift space, including quadratic and tidal galaxy bias. Starting from the corresponding velocity-divergence bispectrum, we derive predictions in both Fourier and configuration space through a spherical-harmonic expansion. We test the Fourier-space calculation against a closed-form expression and assess the numerical convergence of the configuration-space predictions. For a fiducial model, the signal is dominated by a dipolar dependence on triangle orientation relative to the line of sight, while subdominant multipoles distinguish the responses to growth and galaxy bias. In configuration space, the baryon acoustic oscillation (BAO) feature depends on which triangle side crosses the acoustic scale and on the velocity projection. We also relate the model to the density-weighted statistic measured by peculiar-velocity catalogues and formulate estimators for its observable multipoles. This framework supports future measurements and covariance analyses needed to establish the additional cosmological information carried by velocity-density three-point correlations.}

\end{abstract}

\maketitle

\section{\label{sec:intro}Introduction}

\textcolor{darkblue}{Galaxy redshift surveys trace the large-scale distribution of galaxies, providing measurements of the expansion history and growth of cosmic structure. Most constraints have come from two-point statistics, with baryon acoustic oscillations providing a standard ruler and redshift-space distortions constraining the growth rate \citep{Kaiser1987,AlamEtAl2017}. Nonlinear gravitational evolution generates non-Gaussianity whose leading signature in the correlation statistics is the connected three-point correlation function, or its Fourier counterpart, the bispectrum \citep{Fry1984,BernardeauEtAl2002}. Its dependence on triangle shape and orientation helps distinguish gravitational evolution from galaxy bias and provides information beyond two-point correlations \citep{FryGaztanaga1993,VerdeEtAl1998,SugiyamaEtAl2019}.}

\textcolor{darkblue}{Joint two- and three-point analyses have consequently become an important part of spectroscopic-survey cosmology \citep{VerdeEtAl2002,GilMarinEtAl2014,GilMarinEtAl2017,NovellMasotEtAl2025,GuidiEtAl2026}, supported by advances in perturbative modelling and efficient estimators in both Fourier and configuration space \citep{SlepianEisenstein2015,SlepianEisenstein2016,SugiyamaEtAl2021,GuidiEtAl2023,EuclidGuidiEtAl2025}.
In DESI DR1, joint analyses tighten the constraint on the amplitude of fluctuations by approximately $20\%$ with respect to the power spectrum alone \citep{NovellMasotEtAl2025,NovellMasotEtAl2026,ForeroSanchezEtAl2026,GuandalinEtAl2026,ChudaykinEtAl2025}. In configuration space, these estimators have led to measurements of the galaxy three-point correaltion function (3PCF) in several spectroscopic samples \citep{MorescoEtAl2017,VeropalumboEtAl2021,VeropalumboEtAl2022}, to the detection of the baryonic acustic oscillations (BAO) feature in the Baryon Oscillation Spectroscopic Survey (BOSS) \citep{SlepianEtAl2017,MorescoEtAl2021} and the Dark Energy Spectroscopic Instrument (DESI) DR1 \citep{KamalinejadEtAl2026}, and to constraints on massive neutrinos \citep{LabateEtAl2025}. The two representations are complementary, since Fourier space offers a compact model and nearly diagonal covariances, whereas configuration space confines the \textcolor{darkblue}{baryon acoustic oscillation  feature} to a narrow range of separations and simplifies the treatment of the survey window.
\textcolor{darkgreen}{All these measurements, however, concern the galaxy density alone. Extending three-point analyses to peculiar velocities and their correlations with galaxy density is the natural next step.}}

\textcolor{darkblue}{Peculiar velocities provide a dynamical probe of the matter distribution that is independent of the galaxy-density bias relation. By the equivalence principle, galaxies and dark matter respond to the same gravitational acceleration, so their coherent motions trace the same large-scale gravitational field. For an irrotational growing mode, linear theory gives}
\begin{equation}
\textcolor{darkblue}{\mathbf v(\mathbf k,z) = i\,aHf\,\frac{\mathbf k}{k^2}\,\delta_m(\mathbf k,z)\,,}
\label{eq:intro_vlin}
\end{equation}
\textcolor{darkblue}{where $a$ is the scale factor, $H$ the Hubble parameter, $f=\mathrm{d}\ln D/\mathrm{d}\ln a$ the logarithmic growth rate, and $\delta_m$ the matter density contrast. The velocity is therefore related nonlocally to the density, with an amplitude proportional to $aHf$ at fixed density configuration.
Beyond linear theory the instantaneous density field no longer specifies the velocity field uniquely, since the velocity retains information about the preceding evolution. Expressing the equations of motion in terms of $\ln D$ nevertheless substantially reduces their explicit cosmological dependence \citep{NusserColberg1998}, which motivates an approximate scaling of coherent velocities with $aHf$ at comparable density configurations outside strongly nonlinear regions.
Density-velocity two-point correlations have been modelled extensively \citep{Fisher1995,Ma2011,AdamsBlake2017,Nusser2017}. Joint analyses of density and velocity statistics improve constraints on $f\sigma_8$ through their different parameter dependences and partial cancellation of sample variance \citep{KodaEtAl2014,HowlettEtAl2017,Turner2021,Blake2023}.
Including the density-velocity cross-covariance tightens the constraints on $f\sigma_8$ and $\beta=f/b_1$ by approximately $20\%$ \citep{AdamsBlake2017}, but in linear theory these gains leave a common normalisation degeneracy among $f$, the linear galaxy bias $b_1$, and $\sigma_8$.}

\textcolor{darkblue}{A natural extension is the mixed three-point correlation}
\begin{equation}
\textcolor{darkblue}{\left\langle v(\mathbf x_1)\,\delta_g(\mathbf x_2)\,\delta_g(\mathbf x_3)\right\rangle_c\,,
\qquad v(\mathbf x)=\mathbf v(\mathbf x)\cdot\hat{\mathbf n}(\mathbf x)\,,}
\label{eq:intro_vdd}
\end{equation}
\textcolor{darkblue}{where $\delta_g$ is the galaxy density contrast and $\hat{\mathbf n}$ the line of sight. This statistic measures the coupling of a line-of-sight velocity to a pair of density fluctuations, retaining the triangle shape and orientation. \textcolor{darkgreen}{It vanishes for jointly Gaussian, zero-mean fields, so that for Gaussian initial conditions its leading signal is generated by gravitational evolution and nonlinear bias. Its different dependence on the parameters can be seen in real space,} at fixed power-spectrum shape and background geometry, neglecting velocity bias and the residual cosmological dependence of the perturbative kernels. The gravitational contribution scales as $fb_1^2\sigma_8^4$, compared with $fb_1\sigma_8^2$ for the two-point cross-correlation, while quadratic and tidal bias contribute terms proportional to $fb_1b_2\sigma_8^4$ and $fb_1\bG\sigma_8^4$ with different configuration dependences. \textcolor{darkgreen}{Combined with galaxy clustering, these scalings suggest a joint analysis able to separate growth, bias and fluctuation amplitude. The achievable gains, however, depend on the full covariance, on redshift-space effects, on scale cuts and on modelling accuracy.}}

\textcolor{darkblue}{Growing velocity samples strengthen this motivation. DESI Fundamental Plane and Tully--Fisher catalogues \citep{DESIRoss26,DESIDouglass26}, together with Type Ia supernova samples \citep{BoruahEtAl2020,RigaultEtAl2025,CarreresEtAl2023}, expand the scope for distance-based velocity measurements despite their substantial individual uncertainties. At higher redshift, the kinetic Sunyaev--Zel'dovich (kSZ) effect enables velocity reconstruction by combining cosmic microwave background maps with galaxy surveys \citep{SmithEtAl2018,McCarthyEtAl2025}. Future spectroscopic facilities such as the Wide-field Spectroscopic Telescope \citep{MainieriEtAl2024} could provide dense galaxy samples for mixed clustering analyses with these reconstructed fields. Such applications require additional modelling of the reconstruction response, noise and correlations with the density tracers.}

\textcolor{darkblue}{A complete description of the angular structure of the mixed statistic is needed to exploit these opportunities. Existing kSZ bispectrum treatments focus on disconnected products of two-point spectra \citep{SmithEtAl2018}, while the density-density-momentum bispectrum has recently been proposed as a probe of primordial parity violation \citep{YapEtAl2026}. To the best of our knowledge, a model of the anisotropic redshift-space multipoles of $\langle v\delta\delta\rangle$ including galaxy bias is not yet available in either Fourier or configuration space. \textcolor{darkgreen}{Its construction is not a direct extension of the galaxy bispectrum, since the velocity projection} introduces a factor $1/k_1$ which, through triangle closure, couples the directions of both density wavevectors and mixes their angular multipoles.}

\textcolor{darkblue}{In this paper, we develop a tree-level model of this statistic in both spaces. We first construct the bispectrum of the normalised velocity divergence and two galaxy densities, $\theta\delta\delta$, following the approach developed for the galaxy bispectrum by~\citep{SugiyamaEtAl2021}. After factoring out the explicit $aHf$ normalisation, the velocity statistic follows through a geometric map whose mixing tensors are independent of cosmology and can be precomputed. In configuration space, the calculation combines these tensors with double Hankel transforms evaluated using two-dimensional FFTLog \citep{Hamilton2000,Fang2020,GuidiEtAl2023}. We test the Fourier-space implementation against an exact expression and assess the convergence of the angular expansion in both spaces. We then examine the multipole hierarchy, BAO signatures and parameter responses. \textcolor{darkgreen}{Assessing the additional cosmological information carried by the statistic requires a covariance analysis, which is beyond the scope of this work.}}

\textcolor{darkblue}{The model also supplies the new ingredient needed for density-weighted velocity measurements. A catalogue samples the momentum field, $\mathbf p_g=(1+\delta_g)\mathbf v$, rather than a volume-weighted velocity. For Gaussian initial conditions, its three-point correlation with two densities equals $v\delta\delta$ at tree level plus products of density-density and density-velocity two-point correlations, whose models are already available \citep{Blake2023}. We derive this relation and connect the model multipoles to estimators of the catalogue statistic.}

\textcolor{darkblue}{Section~\ref{sec:theory} presents the Fourier- and configuration-space models, their relation to momentum statistics, and their estimators. Section~\ref{sec:pred} examines predictions at $z=0.9$, and Sect.~\ref{sec:conclusions} summarises the results and prospects. The appendices provide the two-point conventions, perturbative kernels and angular derivations, relations between estimator bases, numerical implementation, and convergence tests.}

\section{\label{sec:theory}\rev{Modelling the velocity-density-density three-point function}}

\textcolor{darkblue}{This section develops the tree-level model of the three-point cross-correlation between one peculiar velocity and two galaxy density fields in redshift space. We first compute the bispectrum of the velocity divergence with two densities, project it onto the velocity and decompose it in angular multipoles (Sect.~\ref{sec:model:fourier}). We then transform the multipoles to configuration space (Sect.~\ref{sec:config}), relate them to the momentum statistics sampled by catalogues (Sect.~\ref{sec:theory:momentum}) and construct their estimators (Sect.~\ref{sec:theory:estimator}). The derivations of the multipole formulae and the selection rules are collected in App.~\ref{app:gaunt}.}

\subsection{\label{sec:model:fourier}\rev{Fourier space}}
\subsubsection{\label{sec:theory:fields}\textcolor{darkblue}{Fields and assumptions}}

\textcolor{darkblue}{We work with the galaxy density contrast $\delta_{\rm g}$ and the normalised divergence of the galaxy peculiar-velocity field,}
\begin{equation}
  \theta_{\rm g}(\xv) \equiv -\frac{\vec\nabla\cdot\mathbf{v}_{\rm g}(\xv)}{\Hf} \, ,
  \label{eq:theta_def}
\end{equation}
\textcolor{darkblue}{where $\mathcal{H}=aH$ is the conformal Hubble rate and $f\equiv{\rm d}\ln D/{\rm d}\ln a$ the linear growth rate. Fourier transforms follow the convention $F(\xv)=\int\!\frac{\mathrm{d}^3k}{(2\pi)^3}\,F(\kv)\,e^{i\kv\cdot\xv}$ throughout.}
\textcolor{darkgreen}{The model rests on the following assumptions. The initial conditions are Gaussian, and the dynamics is described by tree-level Eulerian perturbation theory in redshift space \citep{BernardeauEtAl2002}. Velocities are irrotational, since vorticity is not sourced in perturbation theory before shell crossing and its initial value decays with the expansion, and we assume no velocity bias at the order considered. The redshift-space mapping is treated in the plane-parallel approximation, with a fixed line of sight $\nv$. Galaxy bias is truncated at the quadratic order required by tree-level three-point statistics, with the linear bias $b_1$, the quadratic local bias $b_2$ and the tidal bias $\bG$. The two fields then read}

\begin{equation}
\begin{aligned}
\rev{\delta_{\rm g}(\xv)}&\rev{=b_1\,\delta(\xv)+\frac{b_2}{2}\,\delta^2(\xv)+\bG\,\mathcal{G}_2(\Phi_v)(\xv)+\textcolor{darkblue}{\mathcal{O}(\delta^3)}\,,}\\
\rev{\theta_{\rm g}(\xv)}&\rev{=\theta_{\rm m}(\xv)+\mathcal{O}(\nabla^2)\,,}
\end{aligned}
\label{eq:bias_expansion}
\end{equation}
\textcolor{darkblue}{with $\mathcal{G}_2(\Phi_v)=(\partial_i\partial_j\Phi_v)^2-(\nabla^2\Phi_v)^2$, where $\Phi_v$ is the normalised velocity potential, with $\nabla^2\Phi_v=\delta$ at linear order \citep{Desjacques2018}. The second line expresses the absence of velocity bias. By the equivalence principle galaxies and matter respond to the same gravitational acceleration, so a velocity bias can only enter through higher-derivative terms suppressed by $k^2$ \citep{Desjacques2018}, beyond the tree level considered here. In the following we assume Eq.~\eqref{eq:bias_expansion}. Both fields are expanded in standard Eulerian perturbation theory, with redshift-space kernels $Z^{(n)}_\delta$ and $Z^{(n)}_\theta$ that encode gravitational mode coupling, the nonlinear redshift-space mapping and the bias expansion. For the density legs we follow an existing kernel construction \citep{SugiyamaEtAl2021}. The kernels of $\theta_{\rm g}$, which are not given there, are collected with the perturbative expansion in App.~\ref{app:gaunt:kernels}. The parameter set of the model is therefore $\{f,\sigma_8,b_1,b_2,\bG\}$, the same as for the tree-level galaxy bispectrum.}

\subsubsection{\label{sec:theory:tdd}\textcolor{darkblue}{The scalar \texorpdfstring{$\theta\delta\delta$}{theta-delta-delta} bispectrum}}

\textcolor{darkblue}{The basic perturbative object is the cross-bispectrum of the velocity divergence with two galaxy densities. At tree level it receives three contractions, since the second-order field can sit on the divergence leg, $\langle\theta^{(2)}\delta^{(1)}_{\rm g}\delta^{(1)}_{\rm g}\rangle$, or on either density leg, $\langle\theta^{(1)}\delta^{(2)}_{\rm g}\delta^{(1)}_{\rm g}\rangle$ and its image under $\kv_2\leftrightarrow\kv_3$. Only the two density legs are interchangeable, since they sample the same field, while the divergence leg is a distinct field and is not permuted. Both density contractions are included, and no further permutation is applied, since it would count the $\delta^{(2)}_{\rm g}$ term twice. Organising the result by bias monomials gives ten separable amplitude-kernel pairs,}
\begin{align}
  &\langle \theta_{\rm g}(\kv_1)\,\delta_{\rm g}(\kv_2)\,\delta_{\rm g}(\kv_3)\rangle \notag \\
  &\quad= (2\pi)^3\delta_{\rm D}(\kv_1+\kv_2+\kv_3)\,\sum_{p=1}^{10} X^p\,B^p(\kv_2,\kv_3) \, .
  \label{eq:tdd_bisp}
\end{align}
\textcolor{darkblue}{The amplitudes $X^p$ contain the explicit dependence on bias, growth and normalisation. They multiply the bias monomials $b_1^2$, $b_1 f$, $b_1\bG$, $\bG f$, $b_1 b_2$, $b_2 f$, $b_1^2 f$, $b_1 f^2$, $f^2$ and $f^3$, each with the power of $\sigma_8$ required by the tree-level bispectrum. The kernels $B^p(\kv_2,\kv_3)$ contain the dependence on scale and angles at the chosen cosmological template. They are built from the second-order kernels $F^{(2)}_\delta$ and $G^{(2)}_\theta$ and from the redshift-space factors $\mu_i=\hat{\kv}_i\cdot\nv$. Amplitudes and kernels are listed in App.~\ref{app:gaunt:kernels}. The $\theta\delta\delta$ statistic is the building block of the model and carries no bias on the $\theta$ leg. Since it lacks the velocity factor, its multipoles are also free of the multipole mixing described in Sect.~\ref{sec:theory:tripoSH}. It can also be compared directly with data that deliver a velocity-divergence field on a grid, such as velocity fields reconstructed from the density.}

\subsubsection{\label{sec:theory:fourier}\textcolor{darkblue}{From the velocity divergence to the projected velocity}}

\textcolor{darkblue}{For an irrotational field the velocity is fully determined by its divergence, and inverting Eq.~\eqref{eq:theta_def} gives}
\begin{equation}
  \mathbf{v}_{\rm g}(\kv) = i\,\Hf\,\frac{\hat{\kv}}{k}\,\theta_{\rm g}(\kv) \, .
  \label{eq:v_of_theta}
\end{equation}
\textcolor{darkblue}{At linear order $\theta_{\rm g}=\delta_{\rm m}$ and Eq.~\eqref{eq:v_of_theta} reduces to Eq.~\eqref{eq:intro_vlin}. For the velocity component $v\equiv\vv\cdot\mathbf{v}_{\rm g}$ along a direction $\vv$, the bispectrum follows from Eq.~\eqref{eq:tdd_bisp} through the factor $\hat{\kv}_1/k_1$ of the velocity leg,}
\begin{align}
  \textcolor{darkblue}{\langle v(\kv_1)\,\delta_{\rm g}(\kv_2)\,\delta_{\rm g}(\kv_3)\rangle}
  &\textcolor{darkblue}{= (2\pi)^3\delta_{\rm D}(\kv_1+\kv_2+\kv_3)} \notag \\
  &\quad\textcolor{darkblue}{\times\sum_{p=1}^{10} X^p\,\Bt^p(\vv,\kv_2,\kv_3) \, ,} \notag \\
  \textcolor{darkblue}{\Bt^p(\vv,\kv_2,\kv_3)}
  &\textcolor{darkblue}{= i\Hf\,\frac{\hat{\kv}_1\cdot\vv}{k_1}\,B^p(\kv_2,\kv_3) \, .}
  \label{eq:bisp_sugylike}
\end{align}
\textcolor{darkblue}{Peculiar-velocity surveys and kSZ measurements are sensitive to the line-of-sight component only, so the observable is Eq.~\eqref{eq:bisp_sugylike} at $\vv=\nv$. In the multipole decomposition below we nevertheless keep $\vv$ as a formal variable, independent of $\nv$, and set $\vv=\nv$ only when we construct the observable basis in Sect.~\ref{sec:theory:estimator}.}

\textcolor{darkblue}{The entire dependence on the velocity direction resides in the scalar product $\hat{\kv}_1\cdot\vv$, so the bispectrum is exactly dipolar in $\vv$. This holds for any statistic linear in the projected velocity and does not rely on irrotationality. Irrotationality fixes instead the direction of the vector correlator $\langle\mathbf{v}_{\rm g}\delta_{\rm g}\delta_{\rm g}\rangle$, parallel to $\hat{\kv}_1$, and with it the factor $1/k_1$. The dipole in $\vv$ is distinct from the dependence on the line of sight. The line of sight also enters through the redshift-space factors $\mu_i$ of the kernels $B^p$, so the dependence of the observable on the orientation of the triangle relative to $\nv$ contains higher multipoles.}

\textcolor{darkblue}{The factor $1/k_1$ of the velocity leg depends on both density directions, because the constraint $\kv_1+\kv_2+\kv_3=0$ holds for the wavevectors and not for the unit vectors,}
\begin{equation}
  \frac{\hat{\kv}_1\cdot\vv}{k_1} = -\,\frac{k_2\,(\hat{\kv}_2\cdot\vv) + k_3\,(\hat{\kv}_3\cdot\vv)}{k_1^2} \, .
  \label{eq:k1hat_closure}
\end{equation}
\textcolor{darkblue}{The bispectrum therefore admits the exact closed form}
\begin{equation}
  \Bt^p(\vv,\kv_2,\kv_3)
  = -\frac{i\Hf}{k_1^2}\,
    \bigl[k_2\,\mathcal{L}_1(\hat{\kv}_2\cdot\vv)
         +k_3\,\mathcal{L}_1(\hat{\kv}_3\cdot\vv)\bigr]
    B^p \, .
  \label{eq:eq15_closed}
\end{equation}
\textcolor{darkblue}{Equations~\eqref{eq:k1hat_closure} and \eqref{eq:eq15_closed} expose the central difficulty of the velocity statistic. Through the closure, the velocity factor couples the multipoles of the two density legs with the velocity dipole, which is the multipole mixing of Sect.~\ref{sec:theory:tripoSH}. Equation~\eqref{eq:eq15_closed} has two uses. It can be evaluated directly on individual triangles $(\vv,\kv_2,\kv_3)$, without any multipole expansion, and, being exact, it is the reference against which we validate the truncated multipole reconstruction in App.~\ref{app:conv:fourier}.} \rev{The two terms in square brackets map into each other under $\kv_2\leftrightarrow\kv_3$, so the numerator is symmetric in the density legs. The two Gaunt chains of Sect.~\ref{sec:theory:tripoSH} are nevertheless not related by a plain relabelling, because the expansion of $k_1^{-2}$ conjugates the harmonic of $\hat{\kv}_3$ but not that of $\hat{\kv}_2$ (App.~\ref{app:gaunt:bpell}).}

\subsubsection{\label{sec:theory:ir}\textcolor{darkblue}{Infrared resummation}}

\rev{We include the smearing of the BAO feature through infrared resummation. Long-wavelength displacements damp the oscillations, a physical non-linear effect that the tree-level prediction alone does not contain. We split} the linear spectrum into a smooth and a wiggly component, $\Plin = \Pnw + P_{\rm w}$, and exponentiate the large-scale displacement variance on the wiggly part only,
\begin{equation}
  \Plin(k) \;\to\; \Pnw(k) + {\rm e}^{-k^2\Sigma^2(\mu)}\,P_{\rm w}(k) \, ,
  \label{eq:ir_resum}
\end{equation}
with the anisotropic damping $\Sigma^2(\mu)$ including the redshift-space enhancement along the line of sight \citep{Senatore2014, Vlah2016}. \textcolor{darkblue}{We use the form of \citep{IvanovSibiryakov2018}, given with the numerical choices in App.~\ref{app:numerics}. At tree level the kernels are low-order polynomials in $\mu_2$, $\mu_3$ and $\hat{\kv}_2\cdot\hat{\kv}_3$, so their angular expansion would truncate at a finite order. The anisotropic damping is an exponential in $k^2\Sigma^2(\mu)$ and repopulates all higher orders, so the truncated expansion of the next subsection requires a convergence assessment (App.~\ref{app:conv:fourier}).}

\subsubsection{\label{sec:theory:tripoSH}Triple-product spherical-harmonic decomposition}

Being a scalar statistic, $\theta\delta\delta$ involves only the two independent wavevector directions\rev{, while the line of sight $\nv$ enters parametrically through the redshift-space factors $\mu_i$. Its} complete angular content is captured by the double spherical-harmonic decomposition
\begin{align}
  B^p(\kv_2,\kv_3) &= \sum_{\ell_1\ell_2}\sum_{m_1 m_2} B^{p\,m_1 m_2}_{\ell_1\ell_2}(k_2,k_3) \notag \\
  &\quad\times Y^{*m_1}_{\ell_1}(\hat{\kv}_2)\, Y^{m_2}_{\ell_2}(\hat{\kv}_3) \, ,
  \label{eq:tdd_multipoles}
\end{align}
whose coefficients we refer to as the \emph{tables} in what follows. \rev{These tables are the input on which the velocity mixing acts.}

\rev{We decompose $\Bt(\vv,\kv_2,\kv_3)$ on a complete angular basis, the triple-product spherical-harmonic basis. It is a plain product of harmonics of $\hat{\kv}_2$, $\hat{\kv}_3$ and $\vv$, without Clebsch--Gordan coupling into a total angular momentum. The multipoles of each bias kernel are}
\begin{align}
\Bt^{p \, m_3 m_4 m_5}_{\ell_3 \ell_4 \ell_5}(k_2,k_3)
&=
\int d^2 \hat{\kv}_2 \, d^2 \hat{\kv}_3 \, d^2 \vv \,
\Bt^{p}({\vv,\kv_2,\kv_3})
\nonumber\\
&\quad \times
Y^{*m_3}_{\ell_3}(\hat{\kv}_2)\,
Y^{*m_4}_{\ell_4}(\hat{\kv}_3)\,
Y^{*m_5}_{\ell_5}(\vv),
\label{eq:triposh_proj}
\end{align}
\rev{so that, since the spherical harmonics are orthonormal over the full solid angle, the kernel is recovered exactly from its multipoles through the expansion}
\begin{align}
\Bt^{p}({\vv,\kv_2,\kv_3})
&=
\sum_{\ell_3\ell_4\ell_5}
\sum_{m_3m_4m_5}
\Bt^{p\,m_3m_4m_5}_{\ell_3\ell_4\ell_5}(k_2,k_3)
\nonumber\\
&\quad\times
Y^{m_3}_{\ell_3}(\hat{\kv}_2)\,
Y^{m_4}_{\ell_4}(\hat{\kv}_3)\,
Y^{m_5}_{\ell_5}(\vv),
\label{eq:triposh_synth}
\end{align}
with no $(2\ell+1)$ prefactor, since the harmonics are orthonormal. \rev{The projection of Eq.~\eqref{eq:triposh_proj} is also the quantity that an estimator measures, coefficient by coefficient. We construct the estimators in Sect.~\ref{sec:theory:estimator}.}

Were $\hat{\kv}_1$ an independent direction, the projection of Eq.~\eqref{eq:eq15_closed} would be elementary. \rev{The structure of the result comes instead from Eq.~\eqref{eq:k1hat_closure}, where the factor $k_1^{-2}$ must itself be expanded in multipoles of $\hat{\kv}_2\cdot\hat{\kv}_3$. This expansion couples the angular content of the kernels $B^p$, stored in the tables of Eq.~\eqref{eq:tdd_multipoles}, to the output multipoles $(\ell_3,\ell_4)$ through chains of Gaunt integrals.} The derivation, presented step by step in App.~\ref{app:gaunt:bpell}, expresses each velocity multipole as a linear map of the $\theta\delta\delta$ tables,
\begin{align}
\label{eq:bk_p_ell}
  \Bt^{p \, m_3 m_4 m_5}_{\ell_3 \ell_4 \ell_5}(k_2, k_3)
  &= \delta_{1\ell_5} \sum_{\ell_1 \ell_2}\sum_{m_1 m_2}
  \mathcal{M}^{\,m_3 m_4 m_5;\, m_1 m_2}_{\ell_3 \ell_4;\, \ell_1 \ell_2}(k_2,k_3) \notag \\
  &\quad\times B^{p \, m_1 m_2}_{\ell_1 \ell_2}(k_2, k_3) \, ,
\end{align}

\rev{where $\mathcal{M}$ is the \emph{mixing matrix} that maps the $\theta\delta\delta$ tables onto the velocity multipoles at each $(k_2,k_3)$. It is purely geometric, independent of cosmology and of the amplitudes $X^p$, and is built from two chains of three Gaunt integrals weighted by $k_2$, $k_3$ and by the multipole expansion $\Klam(k_2,k_3)$ of $k_1^{-2}$. Its explicit form is given in Eqs.~\eqref{eq:M_explicit} and \eqref{eq:K_Lambda} of App.~\ref{app:gaunt:bpell}. \rev{The sum over the intermediate multipole $\Lambda$ is finite. The Gaunt selection rules bound it by $2\max(\ell^{\rm in}_{\rm max},\ell^{\rm out}_{\rm max})+1$, where $\ell^{\rm in}_{\rm max}$ and $\ell^{\rm out}_{\rm max}$ are the highest multipoles of the $\theta\delta\delta$ tables and of the velocity output. Within this range $\Klam$ scales as $(k_</k_>)^{\Lambda}$, with $k_<$ and $k_>$ the smaller and the larger of $k_2$ and $k_3$, so the high-$\Lambda$ terms matter only for nearly isosceles triangles.}}

The structure of Eq.~\eqref{eq:bk_p_ell} departs in three respects from the familiar $\delta\delta\delta$ case. \rev{First, the Kronecker delta $\delta_{1\ell_5}$ is an exact selection rule, since all velocity-leg multipoles except the dipole vanish at every order in the bias expansion. It follows from the linearity of the statistic in the projected velocity, as noted after Eq.~\eqref{eq:eq15_closed}. In our pipeline the vanishing of the $\ell_5\neq1$ coefficients serves as a check of the numerical projection.}

\rev{Second, the parities of the Gaunt integrals require $\ell_1+\ell_2+\ell_3+\ell_4$ to be odd. In the Newtonian model the kernels are invariant under a joint reflection of $\kv_2$ and $\kv_3$, so only tables with $\ell_1+\ell_2$ even are non-zero and the output has $\ell_3+\ell_4$ odd. Relativistic corrections break this invariance and would populate $\ell_1+\ell_2$ odd, hence $\ell_3+\ell_4$ even. The even-$(\ell_3+\ell_4)$ multipoles therefore provide a null test of the Newtonian model.}

Third, the explicit weights $k_2$ and $k_3$ multiplying the two Gaunt chains are inherited from Eq.~\eqref{eq:eq15_closed}.

\rev{The Fourier-space model thus rests on two objects. The $\theta\delta\delta$ tables of Eq.~\eqref{eq:tdd_multipoles} are computed once per cosmology on a $(k_2,k_3)$ grid, and the mixing operator of Eq.~\eqref{eq:bk_p_ell} is geometric and precomputed. Combined with a radial transform, the same two objects give the configuration-space multipoles of Sect.~\ref{sec:config}.}

\subsection{\label{sec:config}\rev{Configuration space}}

\rev{We now transform the model to configuration space. We first show that a direct evaluation is impractical, so that the multipole decomposition route is necessary (Sect.~\ref{sec:config:structure}), and then derive the radial integrals and the Gaunt mixing (Sect.~\ref{sec:config:radial}).}

\subsubsection{\label{sec:config:structure}Multipole structure}

As in Fourier space, the full three-point function decomposes over the ten bias amplitudes,
\begin{equation}
    \zeta(\vv, \rv_{12}, \rv_{13}) = \sum_{p=1}^{10} X^p \, \zt^{p}(\vv, \rv_{12}, \rv_{13}) \, ,
    \label{eq:zeta_decomp}
\end{equation}
but the two spaces are not on the same computational footing. While $\Bt^p(\vv,\kv_2,\kv_3)$ is algebraic and can be evaluated pointwise through Eq.~\eqref{eq:eq15_closed}, its configuration-space counterpart is defined through the six-dimensional integral
\begin{align}
    \zt^p(\vv,\rv_{12},\rv_{13})\big|_{\rm direct} &= \int \frac{d^3k_2\,d^3k_3}{(2\pi)^6}\,\Bt^p(\vv,\kv_2,\kv_3) \notag \\
    &\quad\times e^{i\kv_2\cdot\rv_{12}}\,e^{i\kv_3\cdot\rv_{13}} \, ,
    \label{eq:zeta_direct}
\end{align}
which is not reducible to one-dimensional radial integrals. \rev{The factor $\hat{\kv}_1\cdot\vv$ inside $\Bt^p$ couples the two angular integrations through Eq.~\eqref{eq:k1hat_closure} and is not separable in $\hat{\kv}_2$ and $\hat{\kv}_3$. A brute-force quadrature of Eq.~\eqref{eq:zeta_direct} on a grid of separations and orientations is therefore prohibitive. The multipole pipeline developed below reduces the problem to two-dimensional radial transforms of precomputed tables, and is the practical route to predictions on a full $(r_{12},r_{13})$ grid.}

This asymmetry between the two spaces also shapes the validation strategy of \rev{App.~\ref{app:convergence}}. \rev{In Fourier space the closed form of Eq.~\eqref{eq:eq15_closed} provides an exact reference for the multipole machinery. In configuration space the multipoles are obtained from the validated Fourier-space objects through radial transforms, whose accuracy we test in App.~\ref{app:numerics}.}

The triple-product structure itself carries over unchanged,
\begin{align}
\zt^{p}(\vv,\rv_{12},\rv_{13})
&=
\sum_{\ell_3\ell_4\ell_5}
\sum_{m_3m_4m_5}
\zeta^{p\,m_3m_4m_5}_{\ell_3\ell_4\ell_5}(r_{12},r_{13})
\nonumber\\
&\quad\times
Y_{\ell_3}^{m_3}(\hat{\rv}_{12})\,
Y_{\ell_4}^{m_4}(\hat{\rv}_{13})\,
Y_{\ell_5}^{m_5}(\vv),
\label{eq:zeta_multipole_structure}
\end{align}
with the multipoles defined, by the orthonormality of the harmonics, as the inverse projection of the full function,
\begin{align}
\zeta^{p\,m_3m_4m_5}_{\ell_3\ell_4\ell_5}(r_{12},r_{13})
&=
\int d^2\hat{\rv}_{12}\, d^2\hat{\rv}_{13}\, d^2\vv \,
\zt^{p}(\vv,\rv_{12},\rv_{13})
\nonumber\\
&\quad\times
Y^{*m_3}_{\ell_3}(\hat{\rv}_{12})\,
Y^{*m_4}_{\ell_4}(\hat{\rv}_{13})\,
Y^{*m_5}_{\ell_5}(\vv),
\label{eq:zeta_config_proj}
\end{align}
the exact configuration-space analogue of the Fourier-space projection of Eq.~\eqref{eq:triposh_proj}.
\rev{Equation~\eqref{eq:zeta_config_proj} also defines the quantity that a potential configuration-space estimator measures (for details, see Sect \ref{sec:theory:estimator}). A velocity-weighted triplet count of the full function of Eq.~\eqref{eq:zeta_direct} becomes expensive for large catalogues, while the harmonic projection lends itself to estimators built from multipole-weighted pair counts or grids, as for the galaxy 3PCF \citep{SlepianEisenstein2015, SlepianEisenstein2018, SugiyamaEtAl2019, FarinaEtAl2024, EuclidVeropalumboEtAl2026}. We construct two such estimators in Sect.~\ref{sec:theory:estimator}.}
The dipolar selection rule $\ell_5=1$ carries over as well. \rev{The plane waves in Eq.~\eqref{eq:zeta_direct} act on $\hat{\kv}_2$ and $\hat{\kv}_3$ but not on the velocity direction, so the transform leaves the $\vv$ dependence untouched. The configuration-space multipoles inherit the exact dipolarity of Sect.~\ref{sec:theory:tripoSH}, and their vanishing for $\ell_5\neq1$ is a check of the numerical pipeline that we report in App.~\ref{app:conv:config}.}

\subsubsection{\label{sec:config:radial}Radial integrals and Gaunt mixing}

The transformation is best introduced through the $\theta\delta\delta$ statistic, whose scalar nature makes it the simplest member of the family. Since Eq.~\eqref{eq:tdd_bisp} contains no factor coupling the two angular integrations, \rev{each Fourier-space multipole of Eq.~\eqref{eq:tdd_multipoles} maps onto a single configuration-space multipole through a two-dimensional Hankel transform, as for the galaxy 3PCF \citep{SugiyamaEtAl2019, GuidiEtAl2023}},
\begin{align}
\zeta^{\,p\,m_1m_2}_{\ell_1\ell_2}(r_{12},r_{13})
&=
\frac{i^{\,\ell_1+\ell_2}}{(2\pi^2)^2}
\int \mathrm{d}k_2\,\mathrm{d}k_3\;
k_2^2\,k_3^2\;
\notag \\ &\quad\times
j_{\ell_1}(k_2r_{12})\,
j_{\ell_2}(k_3r_{13})\,
B^{p\,m_1m_2}_{\ell_1\ell_2}(k_2,k_3),
\label{eq:zeta_tdd}
\end{align}
a strictly diagonal, one-to-one map \rev{without $\Klam$ weight or Gaunt mixing, which preserves both the multipole orders and the azimuthal indices}. Equation~\eqref{eq:zeta_tdd} completes the $\theta\delta\delta$ model of Sect.~\ref{sec:theory:tdd} in configuration space. \rev{The double Hankel transform that appears here is also the elementary operation of the $v\delta\delta$ pipeline. The $\theta\delta\delta$ statistic is therefore the cleanest test of the transform engine (App.~\ref{app:numerics}), and its predictions cost a small fraction of the $v\delta\delta$ ones.}

For $v\delta\delta$, by contrast, the closure factor $\hat{\kv}_1\cdot\vv$ spoils this diagonality. \rev{We expand the plane waves of Eq.~\eqref{eq:zeta_direct} in spherical waves and carry out the angular integrations, with the same Gaunt-chain reduction as in Fourier space. This turns each output multipole} into a linear combination of Hankel transforms of the Fourier-space tables. In terms of the radial transform
\begin{widetext}
\begin{equation}
    \zeta_{r;\ell_1\ell_2\ell_3\ell_4}^{(a)\,p\,m_1m_2}(r_{12}, r_{13}; \Lambda) \equiv -\frac{i\Hf}{3\pi^3} \int \mathrm{d} k_2 \, \mathrm{d} k_3 \, k_2^2 k_3^2 \, k_a \, \Klam(k_2,k_3) \, i^{\ell_3 + \ell_4} j_{\ell_3}(k_2 r_{12}) \, j_{\ell_4}(k_3 r_{13}) \, B_{\ell_1 \ell_2}^{p\,m_1 m_2}(k_2, k_3) \, , \qquad a=2,3 \, ,
    \label{eq:zeta_radial}
\end{equation}
\end{widetext}
which carries the $\Klam$ weight folded inside the double Hankel transform\rev{, together with the weight $k_a$ of the $a$-th closure term}, the configuration-space multipoles take the same mixing-matrix form as in Fourier space,
\begin{widetext}
\begin{equation}
\label{eq:zp_ell}
\zeta_{\ell_3 \ell_4 \ell_5}^{p\,m_3 m_4 m_5}(r_{12},r_{13})
= \delta_{\ell_5,1} \sum_{\ell_1 \ell_2}\sum_{m_1 m_2}\sum_{\Lambda}
{\sum_{a=2,3}
\mathcal{T}^{(a)\,m_3 m_4 m_5;\, m_1 m_2}_{\ell_3 \ell_4;\, \ell_1 \ell_2}(\Lambda)\;
\zeta_{r;\ell_1\ell_2\ell_3\ell_4}^{(a)\,p\,m_1m_2}(r_{12},r_{13};\Lambda)} \, ,
\end{equation}
\end{widetext}
with $\mathcal{T}\rev{^{(a)}}(\Lambda)$ the configuration-space mixing \rev{tensors} and $m_5=-(m_3+m_4)$ fixed by the velocity-leg integral. \rev{Their elements are the geometric counterpart of the Fourier $\mathcal{M}$ of Eq.~\eqref{eq:M_explicit}, with the same two chains of three Gaunt integrals and the same phase $(-1)^{m_1+m_3+m_4}$. 
The two chains correspond to the two closure terms $a=2,3$, whose weights $k_2$ and $k_3$ now sit inside the radial transforms of Eq.~\eqref{eq:zeta_radial}, so that $\mathcal{T}^{(a)}$ is pure geometry. We give the step-by-step derivation in App.~\ref{app:gaunt:zpell}.} \rev{The two mixing matrices differ in one respect only, made explicit by the index $\Lambda$ of $\mathcal{T}$. In Fourier space $\Klam$ multiplies the tables pointwise in $(k_2,k_3)$ inside $\mathcal{M}$. In configuration space it cannot be pulled out of the wavenumber integral and is folded into the radial transform of Eq.~\eqref{eq:zeta_radial}, so that two double Hankel transforms are required for each $\Lambda$ before the contraction with $\mathcal{T}^{(a)}(\Lambda)$.}

\rev{The sum over $\Lambda$ is finite and is not truncated, because the Gaunt selection rules bound it by $\Lambda_{\rm max}=2\max(\ell^{\rm in}_{\rm max},\ell^{\rm out}_{\rm max})+1$ (App.~\ref{app:gaunt}). The only truncation of the pipeline is therefore the output order $\ell^{\rm out}_{\rm max}$, whose convergence we assess in App.~\ref{app:conv:config}.}

The practical pipeline is then a two-step factorisation. \rev{The cosmology-dependent step evaluates the radial transforms of Eq.~\eqref{eq:zeta_radial}, one double Hankel transform per input column, output Bessel pair and closure term $a$. We perform them with the two-dimensional FFTLog algorithm \citep{Hamilton2000, Fang2020}, which integrates the oscillatory Bessel kernels analytically against a power-law decomposition of the integrand (App.~\ref{app:numerics}). The geometric step contracts the results with the precomputed Gaunt-chain tensors of Eq.~\eqref{eq:zp_ell}. The first step runs once for each of the ten channels $p$ and takes between 11 and 27 hours per channel on a 32-core node for our grid of $80\times80$ separation pairs. Predictions for any bias and growth parameters then follow from the stored channels at negligible cost, whereas a direct evaluation of Eq.~\eqref{eq:zeta_direct} on the same grid is out of reach.}

\subsection{\label{sec:theory:momentum}\rev{Momentum statistics}}

\rev{A peculiar-velocity catalogue does not sample the velocity field uniformly in volume. Each velocity is measured at the position of a galaxy, so a sum over the catalogue weights the velocity field by the local number of tracers and measures the momentum $\mathbf p=(1+\delta_{\rm PV})\mathbf v$ \citep{Howlett2019}, where $\delta_{\rm PV}$ is the density contrast of the velocity tracers. For example, the kinetic Sunyaev--Zel'dovich (kSZ) signal is a momentum by construction, with the electron density in place of $\delta_{\rm PV}$. We now relate these statistics to the $v\delta\delta$ bispectrum $B_{v\delta\delta}\equiv\sum_pX^p\Bt^p$ of Eq.~\eqref{eq:bisp_sugylike}, first at three points and then at two.}

\rev{With the momentum on the first leg and $p\equiv\vv\cdot\mathbf p$, the definition of the momentum gives the exact relation
\begin{equation}
\langle p\,\delta_{\rm g}\delta_{\rm g}\rangle=\langle v\,\delta_{\rm g}\delta_{\rm g}\rangle+\langle\delta_{\rm PV}\,v\,\delta_{\rm g}\delta_{\rm g}\rangle\,,
\label{eq:p_dd_exact}
\end{equation}
which holds at all orders and in both Fourier and configuration space. The four-point function splits into products of two-point functions and a connected part. For Gaussian initial conditions the products are of order $\Plin^2$, the same as the tree-level bispectrum, while the connected part is of order $\Plin^3$, the same as the one-loop corrections to $B_{v\delta\delta}$. At tree level, therefore,
\begin{align}
B_{p\delta\delta}(\kv_1,\kv_2,\kv_3)&=B_{v\delta\delta}(\kv_1,\kv_2,\kv_3)+P_{{\rm PV},{\rm g}}(\kv_2)\,P_{v\delta}(-\kv_3)\nonumber\\
&\quad+P_{{\rm PV},{\rm g}}(\kv_3)\,P_{v\delta}(-\kv_2)\,,
\label{eq:bisp_momentum}
\end{align}
where $P_{v\delta}$ is the velocity-galaxy spectrum of Eq.~\eqref{eq:pvd} and $P_{{\rm PV},{\rm g}}$ the cross spectrum of the positions of the two tracers. The third contraction, $\langle\delta_{\rm PV}v\rangle$ at a single point, vanishes by isotropy. In configuration space each product becomes $\xi_{{\rm PV},{\rm g}}(\rv_{12})\,\xi_{v\delta}(\rv_{13})$ or its image, so $\zeta_{v\delta\delta}$ is recovered from the momentum statistic by subtracting two-point functions measured on the same catalogues. The connected term, and the shot noise that appears when velocities and positions come from the same tracers, enter the estimators of Sect.~\ref{sec:theory:estimator}.}

\rev{The same expansion at two points, $\langle p\,\delta_{\rm g}\rangle=\langle v\,\delta_{\rm g}\rangle+\langle\delta_{\rm PV}\,v\,\delta_{\rm g}\rangle$, shows where $B_{v\delta\delta}$ enters the momentum power spectrum. The cross spectrum of momentum and galaxy density is
\begin{equation}
P_{p\delta}(\kv)=P_{v\delta}(\kv)+\int_{\qv}B_{v\delta\delta}(\kv-\qv,\,\qv,\,-\kv)\,,
\label{eq:pk_momentum}
\end{equation}
with the velocity at $\kv-\qv$ and the density of the velocity tracers at $\qv$. In configuration space the integral is the three-point function in the collapsed configuration, in which the velocity sits on one of the two density points and $r_{12}$ or $r_{13}$ vanishes. For dark matter in real space it is the term $A_{01}$ of the momentum-density spectrum $P_{01}$ in the distribution-function approach to redshift-space distortions \citep{Vlah2012}.}

\rev{The two equations use the bispectrum in different ways. In Eq.~\eqref{eq:bisp_momentum} it is a tree-level term evaluated on finite triangles. In Eq.~\eqref{eq:pk_momentum} it is integrated over all the triangles that close on $\kv$, at the same order in $\Plin$ as the loop terms of Eq.~\eqref{eq:loop_pk}. The integral contains terms proportional to the linear velocity dispersion $\sigma_v^2$ and to the variance $\sigma_0^2$ of the linear field \citep{Vlah2012}, so it depends on all scales and is not a separate observable.}

\rev{The density-weighted pairwise velocity has the same structure. Its definition, $[1+\xi(r)]\,v_{12}(r)=\langle(1+\delta_1)(1+\delta_2)(\mathbf v_1-\mathbf v_2)\cdot\hat{\rv}\rangle$, contains at second order the collapsed function $\langle\delta_1\delta_2\,\mathbf v_1\cdot\hat{\rv}\rangle$ twice. This term is part of the second-order solution of the pair conservation equation \citep{JuszkiewiczSpringelDurrer1999}. We conclude that the momentum-density-density bispectrum is the lowest-order statistic in which the connected $v\delta\delta$ term appears on finite triangles.}

\rev{For the kSZ effect the picture is similar. The kSZ temperature shift of a galaxy is proportional to the line-of-sight momentum of the electrons around it \citep{HandEtAl2012}. The galaxy-galaxy-temperature bispectrum of kSZ tomography is a projected momentum-density-density bispectrum, with the electron momentum $\delta_{\rm e}v_r$ in place of $p$, and its tree-level model keeps only the disconnected product of the large-scale galaxy-velocity spectrum and the small-scale galaxy-electron spectrum \citep{SmithEtAl2018}. This is the kSZ counterpart of the two products of Eq.~\eqref{eq:bisp_momentum}, and it dominates in the squeezed limit. The pairwise kSZ signal \citep{HandEtAl2012} measures the density-weighted pairwise velocity and contains $v\delta\delta$ only in the collapsed form discussed above.}

\rev{The connected term becomes accessible when the large-scale velocity reconstructed from the kSZ effect is correlated with two galaxy fields. To the extent that the reconstruction traces the large-scale velocity, this measures $\zeta_{v\delta\delta}$ on the scales of the reconstruction, up to the optical-depth degeneracy of its amplitude \citep{SmithEtAl2018}. In the rest of this paper we model the connected term alone, since it is the new ingredient common to all these statistics.}

\subsection{\label{sec:theory:estimator}\rev{Estimators}}

\textcolor{darkgreen}{A velocity catalogue provides one line-of-sight velocity per galaxy, and the statistic is built by pairing it with the density of the surrounding galaxies. An estimator therefore does not measure the full set of triple-product coefficients of Eq.~\eqref{eq:zeta_config_proj}, since only the line-of-sight component of the velocity is observed and the observable is the slice $\vv=\nv$.} \rev{ Taking the line of sight as the polar axis, only the coefficients with $m_5=0$ and $m_4=-m_3$ survive, and we write them as $c_{\ell_3\ell_4m}$ with $m\equiv m_3$. They define the Slepian--Eisenstein (SE) basis, which we derive from Eq.~\eqref{eq:zp_ell} in App.~\ref{app:basis}. The same coefficients can be coupled into a multipole $L$ of the dependence on the line of sight,}
\begin{equation}
\label{eq:Lbasis}
\rev{S_{\ell_3\ell_4L}
= \sqrt{4\pi(2L+1)}\, \sum_{m} (-1)^{m}
\begin{pmatrix} \ell_3 & \ell_4 & L \\ m & -m & 0 \end{pmatrix}
c_{\ell_3\ell_4 m}\,,}
\end{equation}
\rev{which is the same unitary change of basis as in Refs.~\cite{SugiyamaEtAl2019,SugiyamaEtAl2021}, without their normalisation factor. For $v\delta\delta$ the selection rules of App.~\ref{app:gaunt:rules} allow only odd $L$ with $\ell_3+\ell_4$ odd. The term $L=1$ contains the projection $\vv\cdot\nv$ of the real-space dipole, while $L\geq3$ is generated only by redshift-space distortions of the density legs. For $\theta\delta\delta$ the same construction gives even $L$ and even $\ell_3+\ell_4$.}

\textcolor{darkblue}{The model predicts these coefficients directly. With $\nv$ as the polar axis, Eqs.~\eqref{eq:Lbasis} and \eqref{eq:se_coeff} give for $v\delta\delta$}
\begin{equation}
\label{eq:S_model}
\textcolor{darkblue}{\begin{aligned}S_{\ell_3\ell_4L}&=\sqrt{3(2L+1)}\sum_{m}
\begin{pmatrix} \ell_3 & \ell_4 & L \\ m & -m & 0 \end{pmatrix}\\
&\quad\times\sum_{p}X^p\,\mathcal{F}^{\,p\,m,-m,0}_{\ell_3\ell_4 1}\,,\end{aligned}}
\end{equation}
\textcolor{darkblue}{where $\mathcal{F}$ stands for the multipoles $\Bt$ of Eq.~\eqref{eq:bk_p_ell} as functions of $(k_2,k_3)$, or for the multipoles $\zeta$ of Eq.~\eqref{eq:zp_ell} as functions of $(r_{12},r_{13})$. For $\theta\delta\delta$ the azimuthal symmetry around $\nv$ leaves only the tables with $m_1=m_2$, and the same rotation gives}
\begin{equation}
\label{eq:S_model_tdd}
\textcolor{darkblue}{\begin{aligned}S_{\ell_3\ell_4L}&=\sqrt{4\pi(2L+1)}\sum_{m}(-1)^m
\begin{pmatrix} \ell_3 & \ell_4 & L \\ m & -m & 0 \end{pmatrix}\\
&\quad\times\sum_{p}X^p\,\mathcal{F}^{\,p\,mm}_{\ell_3\ell_4}\,,\end{aligned}}
\end{equation}
\textcolor{darkblue}{with $\mathcal{F}$ the tables $B$ of Eq.~\eqref{eq:tdd_multipoles} or their transforms $\zeta$ of Eq.~\eqref{eq:zeta_tdd}. We use the same symbol $S_{\ell_3\ell_4L}$ in both spaces. These are the coefficients shown in Sect.~\ref{sec:pred}, and the quantities targeted by the estimators below.}

\rev{The $L$ basis corresponds to the rotation-invariant functions
\begin{equation}
\begin{split}
\mathcal{Y}_{\ell_3\ell_4L}(\hat{\rv}_{12},\hat{\rv}_{13},\nv)&=\sum_{m_3m_4M}\begin{pmatrix}\ell_3&\ell_4&L\\m_3&m_4&M\end{pmatrix}\\
&\quad\times Y^{m_3}_{\ell_3}(\hat{\rv}_{12})\,Y^{m_4}_{\ell_4}(\hat{\rv}_{13})\,Y^{M}_{L}(\nv)\,,
\end{split}
\label{eq:Ybasis}
\end{equation}
which are orthonormal on the three spheres and give $\zeta=\sum S_{\ell_3\ell_4L}\,\mathcal{Y}_{\ell_3\ell_4L}$ for any orientation of $\nv$. We place the velocity on galaxy~1, with $u(\xv)=\mathbf v_{\rm g}(\xv)\cdot\hat{\xv}$ measured along its own line of sight, and the densities on galaxies 2 and 3 at $\xv+\rv_{12}$ and $\xv+\rv_{13}$. The estimator
\begin{equation}
\begin{split}
\hat S_{\ell_3\ell_4L}(r_{12},r_{13})&=\frac{4\pi}{V}\int\!\mathrm{d}^3x\,u(\xv)\int\!\mathrm{d}\Omega_{12}\,\mathrm{d}\Omega_{13}\,\delta_{\rm g}(\xv+\rv_{12})\\
&\quad\times\delta_{\rm g}(\xv+\rv_{13})\,\mathcal{Y}^*_{\ell_3\ell_4L}(\hat{\rv}_{12},\hat{\rv}_{13},\hat{\xv})\,,
\end{split}
\label{eq:L_estimator}
\end{equation}
with radial binning implied, is unbiased in the plane-parallel limit. At fixed line of sight the angular integral returns $S_{\ell_3\ell_4L}/4\pi$ for every orientation of $\hat{\xv}$, so the average over positions replaces the integral over $\nv$.}

\rev{The basis function of Eq.~\eqref{eq:Ybasis} factorises into three harmonics. Equation~\eqref{eq:L_estimator} is then a sum of products of the velocity weighted by $Y^{M*}_L(\hat{\xv})$ with two densities convolved with spherical-shell kernels, which can all be computed with fast Fourier transforms as for the anisotropic bispectrum \citep{SugiyamaEtAl2019}. The Fourier-space estimator \textcolor{darkblue}{$\hat S_{\ell_3\ell_4L}(k_2,k_3)$} follows with $\hat{\rv}\to\hat{\kv}$. This global-coordinate construction suits large density samples and Fourier space.}

\rev{The SE coefficients are instead the natural target of the pair-counting algorithm of Refs.~\citep{SlepianEisenstein2015,SlepianEisenstein2018}. \textcolor{newgreen}{\citep{FarinaEtAl2024} presented also an independent implementation in the code \texttt{MeasCorr} which represents the baseline for future modifications. The same algorithm is also adopted in the \textit{Euclid} data-analysis pipeline, as described by~\citep{EuclidVeropalumboEtAl2026}.} Around each primary position $\xv$ one computes the harmonic coefficients of the density in a radial bin, $a_{\ell m}(r;\xv)=\int\mathrm{d}\Omega_{\hat{\rv}}\,\delta_{\rm g}(\xv+\rv)\,Y^{m\,*}_{\ell}(\hat{\rv})$, in a frame rotated so that the local line of sight is the polar axis. Weighting the product of two such coefficients by the velocity of the primary gives}
\begin{equation}
\rev{\begin{aligned}\hat{c}_{\ell_3\ell_4m}(r_{12},r_{13})
&=\frac{1}{V}\int\mathrm{d}^3x\;u(\xv) \\&\quad\times a_{\ell_3m}(r_{12};\xv)\,a^{*}_{\ell_4m}(r_{13};\xv)\,,\end{aligned}}
\label{eq:se_estimator}
\end{equation}
\rev{whose expectation value is $c_{\ell_3\ell_4m}$ by the orthonormality of the spherical harmonics. The azimuthal orientation of the rotated frame is irrelevant, since the phase $e^{im\varphi_0}$ cancels in the product. With the velocity on the primaries only, the cost scales as the number of velocity tracers times the number of their neighbours, which is small for peculiar-velocity catalogues. The output coincides with the model coefficients of Eq.~\eqref{eq:se_coeff}, with no change of basis.}

\rev{Both estimators also measure the parity-odd modes, at no extra cost. Under a parity transformation $\hat{\rv}_{12}$, $\hat{\rv}_{13}$ and $\nv$ reverse together, while $u$ is a true scalar because the velocity is a polar vector. The basis function of Eq.~\eqref{eq:Ybasis} then acquires the sign $(-1)^{\ell_3+\ell_4+L}$, so a parity-invariant model predicts $S_{\ell_3\ell_4L}=0$ for $\ell_3+\ell_4+L$ odd. In the SE basis these modes are the imaginary parts of $c_{\ell_3\ell_4m}$ (App.~\ref{app:basis}), which vanish in our model to numerical precision.}

\rev{On a catalogue the integral over $\xv$ becomes a sum over the velocity tracers, so both estimators measure the momentum statistic of Eq.~\eqref{eq:p_dd_exact} rather than $v\delta\delta$. At tree level the difference is the pair of products of Eq.~\eqref{eq:bisp_momentum}. We define $\xi_{v\delta}(\rv)\equiv\langle u(\xv)\,\delta_{\rm g}(\xv+\rv)\rangle$ and $\xi_{{\rm PV},{\rm g}}(\rv)\equiv\langle\delta_{\rm PV}(\xv)\,\delta_{\rm g}(\xv+\rv)\rangle$, with Legendre multipoles $\xi^\ell$ in $\hat{\rv}\cdot\nv$. The products then project onto the $L$ basis as}
\begin{equation}
\rev{\begin{aligned}
S^{\rm disc}_{\ell_3\ell_4L}&=(4\pi)^{3/2}\sqrt{\frac{2L+1}{(2\ell_3+1)(2\ell_4+1)}}\begin{pmatrix}\ell_3&\ell_4&L\\0&0&0\end{pmatrix}\\
&\quad\times\Big[\xi^{\ell_3}_{{\rm PV},{\rm g}}(r_{12})\,\xi^{\ell_4}_{v\delta}(r_{13})+\xi^{\ell_3}_{v\delta}(r_{12})\,\xi^{\ell_4}_{{\rm PV},{\rm g}}(r_{13})\Big]\,,
\end{aligned}}
\label{eq:S_disc}
\end{equation}
\rev{so that $\langle\hat S_{\ell_3\ell_4L}\rangle=S_{\ell_3\ell_4L}+S^{\rm disc}_{\ell_3\ell_4L}$ at tree level. The products are separable in $(r_{12},r_{13})$ and fall in the same parity class as the signal, with $\ell_3+\ell_4$ and $L$ odd. They must therefore be subtracted with two-point functions measured on the same catalogues, and their SE counterparts follow from Eq.~\eqref{eq:se_from_L}.}

\rev{Volume-weighted velocities avoid this subtraction. They can be obtained from a Voronoi tessellation of the velocity tracers \citep{BernardeauWeygaert1996} or from a reconstructed velocity field, at the price of modelling the sampling and the smoothing. Beyond tree level the connected part of Eq.~\eqref{eq:p_dd_exact} enters together with the one-loop corrections to $B_{v\delta\delta}$. When the velocity tracers also belong to the density sample, coincident points add shot-noise terms to the estimators. We defer these terms, the survey window, the velocity noise and the wide-angle corrections to the local line of sight to future work, together with the covariance.}

\section{\label{sec:pred}\rev{Predictions}}

\begin{figure}[!t]
\centering
\includegraphics[width=\columnwidth]{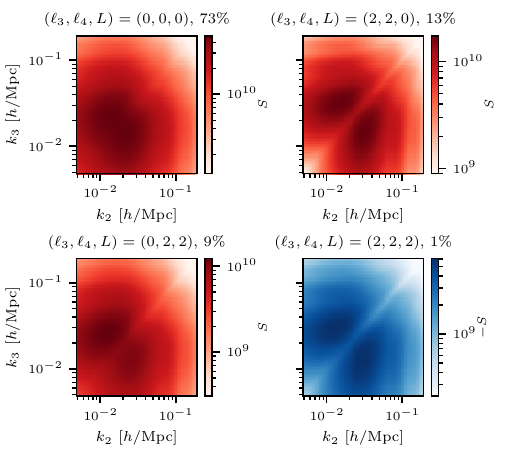}
\caption{\label{fig:tdd_fourier_signal}\rev{Leading coefficients $S_{\ell_3\ell_4L}(k_2,k_3)$ of the $\theta\delta\delta$ bispectrum for the fiducial parameters. Each coefficient has a single sign and is shown on a logarithmic colour scale that spans its own range, from the first percentile to the maximum of $|S_{\ell_3\ell_4L}|$. The negative coefficient $(2,2,2)$ is shown in blue as $-S_{\ell_3\ell_4L}$. The fraction of the total amplitude carried by each coefficient is given in the panel title.}}
\end{figure}

\begin{figure}[!t]
\centering
\includegraphics[width=\columnwidth]{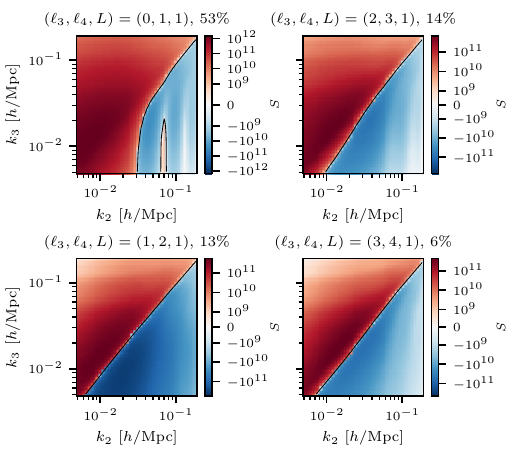}
\caption{\label{fig:vdd_fourier_signal}\rev{Same as Fig.~\ref{fig:tdd_fourier_signal} for the $v\delta\delta$ bispectrum, in units of $i\mathcal{H}$. The coefficients change sign and are shown on a symmetric logarithmic scale, which covers the decades populated by each coefficient, at most four, and is linear below them. Black lines mark the zeros.}}
\end{figure}

\textcolor{darkblue}{\textcolor{darkgreen}{This section describes what the model predicts for the velocity-density-density correlation, in Fourier and in configuration space, for a common fiducial model. The velocity statistic is dominated by its lowest line-of-sight sector (Figs.~\ref{fig:tdd_fourier_signal} and \ref{fig:vdd_fourier_signal}). In Fourier space its signal} lives on larger scales than that of the scalar $\theta\delta\delta$, and the sign of most coefficients tells which of the two density sides is shorter. The growth rate is separated from the overall amplitude only through the ratio between line-of-sight sectors, while the second-order biases change the shape of the signal but \textcolor{darkgreen}{remain} degenerate with each other. \textcolor{darkgreen}{Finally, the} BAO signal of a single triangle depends on which side crosses the acoustic scale and on the direction of the velocity.}

\textcolor{darkblue}{The fiducial parameters are listed in Table~\ref{tab:fiducial}, at the reference redshift $z=0.9$. The linear power spectrum has the shape of the Planck 2018 spectrum, and its no-wiggle counterpart enters the infrared resummation of Eq.~\eqref{eq:ir_resum}. The amplitude of the spectrum enters only through $\sigma_8$ in the amplitudes $X^p$. We take $f$ and $\sigma_8$ from Planck 2018 at $z=0.9$. For a linear bias $b_1=2$, the quadratic bias follows the fit to simulations of \citep{LazeyrasEtAl2016} and the tidal bias the local Lagrangian relation. App.~\ref{app:numerics} collects the numerical settings.} 

The redshift $z = 0.9$ is for illustrative purposes. Distance-based velocity catalogues lie
mostly at $z \lesssim 0.1$, where the plane-parallel and tree-level
approximations adopted here are least accurate and wide-angle corrections become
necessary. At higher redshift, kSZ velocity reconstruction is already available
at $z \sim 0.4$--$1$~\cite{McCarthyEtAl2025}, and WST~\cite{MainieriEtAl2024} will
provide dense galaxy samples. The model is not tied to this choice: the redshift
enters only through $f$, $\sigma_8$ and the biases in the amplitudes $X^p$,
through $\mathcal{H}$, and through the damping of
Eq.~\eqref{eq:ir_resum}, so the same channels yield predictions at any
redshift once the infrared resummation is recomputed.

We show the observable coefficients $S_{\ell_3\ell_4L}$ of Sect.~\ref{sec:theory:estimator} in the $L$ basis of Eq.~\eqref{eq:Lbasis}, and their Slepian--Eisenstein counterparts in App.~\ref{app:basis}. The predictions concern the velocity statistic. The momentum statistic sampled by a catalogue contains in addition the products of two-point functions derived in Sects.~\ref{sec:theory:momentum} and \ref{sec:theory:estimator}.

\begin{table}
\caption{\label{tab:fiducial}\rev{Fiducial parameters of the predictions at $z=0.9$. The damping scales are those of Eqs.~\eqref{eq:ir_resum} and \eqref{eq:sigma2_mu}.} \textcolor{darkblue}{They exceed by $4\%$ and $8\%$ the values computed from the Planck spectrum at $z=0.9$, a difference far below the effects discussed in Sect.~\ref{sec:pred}.}}
\begin{ruledtabular}
\begin{tabular}{lcl}
\rev{Parameter} & \rev{Value} & \rev{Origin}\\
\hline
\rev{$b_1$} & \rev{$2.0$} & \rev{chosen}\\
\rev{$b_2$} & \rev{$-0.094$} & \rev{fit of \citep{LazeyrasEtAl2016} at $b_1=2$}\\
\rev{$\bG$} & \rev{$-2/7$} & \rev{local Lagrangian, $-2(b_1-1)/7$}\\
\rev{$f$} & \rev{$0.86$} & \rev{Planck 2018}\\
\rev{$\sigma_8$} & \rev{$0.51$} & \rev{Planck 2018}\\
\rev{$\Sigma^2$} & \rev{$12.0\,(h^{-1}{\rm Mpc})^2$} & \rev{IR resummation}\\
\rev{$\delta\Sigma^2$} & \rev{$3.3\,(h^{-1}{\rm Mpc})^2$} & \rev{IR resummation}\\
\end{tabular}
\end{ruledtabular}
\end{table}

\begin{table*}
\caption{\label{tab:ranking_fourier}\rev{Fraction of the amplitude of Eq.~\eqref{eq:Pamp} carried by the leading Fourier-space coefficients, over $0.005\le k_{2,3}\le0.2\hMpc$, for the signal, the BAO difference and the derivatives with respect to $\ln f$, $b_2$ and $\bG$. Each row pairs a $\theta\delta\delta$ coefficient with its main image under the velocity projection, and each entry reads $\theta\delta\delta/v\delta\delta$. The main channel carries between $26\%$ and $61\%$ of the $\theta\delta\delta$ coefficient, and the rest flows into neighbouring cells. Fractions include the multiplicity of the coefficient. The last two rows give the totals of the lowest and of the next $L$ sector.}}
\begin{ruledtabular}
\begin{tabular}{lccccc}
$\theta\delta\delta\to v\delta\delta$ & signal & BAO & $\partial_{\ln f}$ & $\partial_{b_2}$ & $\partial_{\bG}$ \\
\hline
$(0,0,0)\to(0,1,1)$ & 0.73/0.53 & 0.39/0.41 & 0.34/0.47 & 0.94/0.63 & 0.71/0.35 \\
$(1,1,0)\to(1,2,1)$ & 0.01/0.13 & 0.26/0.18 & 0.01/0.14 & 0.05/0.20 & 0.20/0.37 \\
$(2,2,0)\to(2,3,1)$ & 0.13/0.14 & 0.14/0.16 & 0.13/0.14 & 0.01/0.08 & 0.05/0.13 \\
$(0,2,2)\to(0,3,3)$ & 0.09/0.02 & 0.05/0.01 & 0.37/0.04 & 0.01/$<$0.01 & 0.01/$<$0.01 \\
$L=0\to L=1$ total & 0.88/0.94 & 0.87/0.93 & 0.48/0.87 & 0.99/$>$0.99 & 0.99/$>$0.99 \\
$L=2\to L=3$ total & 0.12/0.06 & 0.13/0.07 & 0.50/0.12 & 0.01/$<$0.01 & 0.01/$<$0.01 \\
\end{tabular}
\end{ruledtabular}
\end{table*}

\begin{figure*}
\centering
\includegraphics[width=0.9\textwidth]{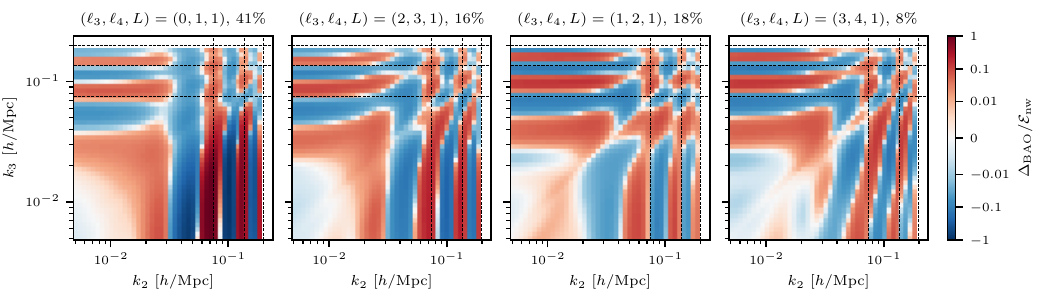}
\caption{\label{fig:vdd_fourier_bao}\rev{Difference $\Delta_{\rm BAO}$ between the $v\delta\delta$ coefficients computed with the full and with the no-wiggle linear power spectrum, divided by the smoothed envelope $\mathcal{E}_{\rm nw}$ of the no-wiggle coefficient \textcolor{darkblue}{(Eq.~\eqref{eq:envelope})}, for the same four coefficients as in Fig.~\ref{fig:vdd_fourier_signal}. The colour scale is linear within $\pm1\%$ and logarithmic beyond. Dashed lines mark the first three acoustic peaks of the linear spectrum, $k=0.075$, $0.137$ and $0.199\hMpc$. The percentages give the share of the amplitude of $\Delta_{\rm BAO}$.}}
\end{figure*}

\textcolor{darkblue}{Both bases are orthonormal, so the total power is shared among the coefficients without loss. We rank the coefficients by their squared amplitude over the plotted range,}
\begin{equation}
\label{eq:Pamp}
\rev{\mathcal{P}_{\ell_3\ell_4L}
= \int_{x_{\rm min}}^{x_{\rm max}} \! dx_2 \int_{x_{\rm min}}^{x_{\rm max}} \! dx_3\, \big| S_{\ell_3\ell_4L}(x_2,x_3) \big|^2\,,}
\end{equation}
\textcolor{darkblue}{where $x$ stands for $k$ in Fourier space and for $r$ in configuration space, with $0.005\le k\le0.2\hMpc$ and $20\le r\le180\Mpc$. The two orderings $(\ell_3,\ell_4)$ and $(\ell_4,\ell_3)$ carry the same amplitude, and we count them once with multiplicity two. We quote each $\mathcal{P}$ as a fraction of the sum over all computed coefficients, up to $\ell_3,\ell_4\le8$ for $\theta\delta\delta$ and $\ell_3,\ell_4\le14$ for $v\delta\delta$ (App.~\ref{app:convergence}). These fractions describe how the amplitude is distributed within the chosen ranges and multipole set. They are not a measure of information, which requires the covariance of the estimator (App.~\ref{app:conv:metric}).}

\subsection{\label{sec:pred:fourier}\rev{Fourier space}}

\subsubsection{\label{sec:pred:fourier:signal}\textcolor{darkblue}{Signal hierarchy and scale dependence}}

\begin{figure*}
\centering
\includegraphics[width=0.85\textwidth]{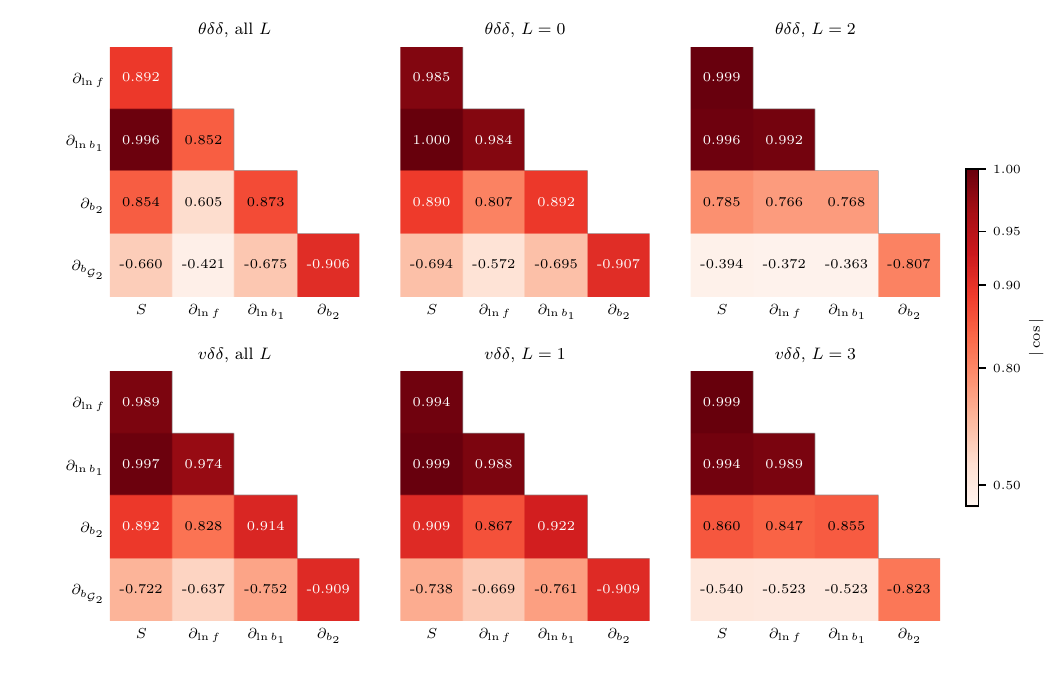}
\caption{\label{fig:similarity_fourier}\rev{Cosine similarity, Eq.~\eqref{eq:cosine}, between the signal $S$ and its derivatives with respect to $\ln f$, $\ln b_1$, $b_2$ and $\bG$, for $\theta\delta\delta$ (top) and $v\delta\delta$ (bottom). The left column combines all $L$, the central and right columns use the lowest and the next $L$ sector only. The colour encodes $|\cos|$ on a stretched scale that separates values close to unity, and the signed value is printed in each cell.}}
\end{figure*}

\textcolor{darkblue}{The factor $\hat{\kv}_1\cdot\nv$ of the velocity is a dipole, so the projection moves the signal from the lowest even line-of-sight sector to the lowest odd one. Figure~\ref{fig:tdd_fourier_signal} shows the four leading coefficients of the scalar $\theta\delta\delta$ as functions of $k_2$ and $k_3$, and Fig.~\ref{fig:vdd_fourier_signal} those of $v\delta\delta$. The sector $L=0$ carries $88\%$ of the amplitude of $\theta\delta\delta$, defined in Eq.~\eqref{eq:Pamp}, and the sector $L=1$ carries $94\%$ of that of $v\delta\delta$. In both correlators the leading coefficient is the monopole of the density legs, $(0,0,0)$ and its image $(0,1,1)$. Table~\ref{tab:ranking_fourier} pairs each $\theta\delta\delta$ coefficient with its main image under the projection and gives the individual shares. The pairs mark the dominant route of the mixing of Sect.~\ref{sec:theory:tripoSH}, not a one-to-one correspondence.}

\textcolor{darkblue}{The velocity also carries the factor $1/k_1$ of Eq.~\eqref{eq:v_of_theta}, which weights the signal towards large scales. More than $80\%$ of the amplitude of each $v\delta\delta$ coefficient lies at $k_2,k_3<0.05\hMpc$, against about half for $\theta\delta\delta$.}

\textcolor{darkblue}{The same factor explains the sign of the coefficients. At fixed $k_2$ and $k_3$ it is largest when the two density wavevectors are nearly antiparallel. There $k_1=|k_2-k_3|$, and $\kv_1=-(\kv_2+\kv_3)$ points along or against $\kv_3$ depending on which of the two sides is longer. The coefficients with $\ell_3\neq\ell_4$ therefore change sign close to the diagonal $k_2=k_3$. They are mostly positive for $k_3>k_2$ and negative for $k_3<k_2$, so their sign tells which of the two density wavevectors is shorter.}

\subsubsection{\label{sec:pred:fourier:bao}\textcolor{darkgreen}{Baryon acoustic oscillations in the multipoles}}

\textcolor{darkblue}{We isolate the BAO signal as the difference $\Delta_{\rm BAO}$ between the coefficients computed with the full and with the no-wiggle linear spectrum, shown in Fig.~\ref{fig:vdd_fourier_bao}. To make it visible on top of the steep trend of the signal, we divide it by a smoothed envelope of the no-wiggle coefficient, which stays positive where the coefficient crosses zero (App.~\ref{app:numerics}).}

\textcolor{darkblue}{The oscillations run along lines of constant $k_2$ and constant $k_3$, at the peaks of the linear spectra of the two density legs. The wiggles of $P(k_1)$, the spectrum of the velocity leg, leave no visible pattern, because at fixed $(k_2,k_3)$ the side $k_1$ varies with the opening angle and the multipole projection averages over it. The fractional amplitude reaches $4$ to $5\%$ once one of the two sides exceeds $0.06\hMpc$. We follow the oscillations of all three sides on individual triangles in Sect.~\ref{sec:pred:fourier_triangles}.}

\textcolor{darkblue}{The BAO signal is not shared among the coefficients like the full signal. In $\theta\delta\delta$ the dipole $(1,1,0)$, nearly empty in the signal, carries a quarter of the difference. Table~\ref{tab:ranking_fourier} gives the shares for both correlators, and we show only the $v\delta\delta$ maps.}

\subsubsection{\label{sec:pred:fourier:resp}\textcolor{darkblue}{Response to the parameters}}

\textcolor{darkblue}{We now ask which parameters change the shape of the signal and which only its amplitude. The model is a polynomial in the parameters, Eq.~\eqref{eq:bisp_sugylike}, so we compute the derivatives analytically from the bias-independent channels, with the damping of Eq.~\eqref{eq:ir_resum} held at its fiducial value. At tree level the bispectrum is a product of two linear spectra and scales as $\sigma_8^4$, so $\partial_{\ln\sigma_8}$ is four times the signal and carries no information on the shape. The linear bias enters most terms of Eq.~\eqref{eq:bisp_sugylike}, and $\partial_{\ln b_1}$ is also close to a rescaling of the signal. The growth rate instead moves amplitude towards high $L$, because the redshift-space terms carry higher powers of $f$ and of the line-of-sight cosines. Table~\ref{tab:ranking_fourier} gives the shares of each derivative.}

\textcolor{darkblue}{The shares tell where a derivative lives, but not whether it can be told apart from another. We compare two responses $D_a$ and $D_b$ through their cosine similarity}
\begin{align}
\textcolor{darkblue}{\cos(D_a,D_b)} &\textcolor{darkblue}{= \frac{\langle D_a,D_b\rangle}{\sqrt{\langle D_a,D_a\rangle\langle D_b,D_b\rangle}}\,,} \label{eq:cosine}\\
\textcolor{darkblue}{\langle D_a,D_b\rangle} &\textcolor{darkblue}{= \sum_{\ell_3\ell_4L}\int\!dk_2\int\!dk_3\, D_{a,\ell_3\ell_4L}\,D_{b,\ell_3\ell_4L}\,,} \nonumber
\end{align}
\textcolor{darkblue}{where the sum runs over the coefficients of a given set of $L$ and the integrals over the range of Eq.~\eqref{eq:Pamp}. \textcolor{darkgreen}{We denote by $S$ the set of coefficients $S_{\ell_3\ell_4L}$ of Eqs.~\eqref{eq:S_model} and \eqref{eq:S_model_tdd} evaluated at the fiducial parameters of Table~\ref{tab:fiducial}, which we call the signal, and by $\partial_\theta S$ its derivative with respect to the parameter $\theta$.} A cosine close to unity means that the two responses differ only in amplitude, so that a change in one parameter can be absorbed by the other. The cosine ignores the covariance of the estimator, so it compares shapes and not information content. Figure~\ref{fig:similarity_fourier} shows the cosines between the signal and its four derivatives, for all $L$ together and for the two lowest sectors separately.}

\textcolor{darkblue}{Within a single sector $\partial_{\ln f}$ and $\partial_{\ln b_1}$ are rescalings of the signal. A single sector therefore constrains only the combination of $f$, $b_1$ and $\sigma_8$ that sets its amplitude, and the growth rate is separated from it by the ratio between sectors. This handle is weaker in $v\delta\delta$, whose higher sector carries only $6\%$ of the signal. Combining the sectors lowers the cosine between $\partial_{\ln f}$ and the signal to $0.89$ in $\theta\delta\delta$, but leaves it at $0.99$ in $v\delta\delta$. The weight of the redshift-space sector is what a covariance analysis has to quantify.}

\textcolor{darkblue}{The second-order biases change the shape of the signal, through kernels with a different angular dependence. The quadratic bias multiplies a kernel independent of the angle between the two wavevectors of each term, while the tidal kernel depends on it. The derivative with respect to $\bG$ therefore puts more amplitude than that with respect to $b_2$ into the coefficients with $\ell_3,\ell_4\geq1$, $(1,1,0)$ in $\theta\delta\delta$ and $(1,2,1)$ in $v\delta\delta$. Over all coefficients the two responses remain degenerate, with $|\cos|\simeq0.91$ in both correlators, so what separates them is carried by these higher terms.}

\subsubsection{\label{sec:pred:fourier_triangles}\rev{Equilateral triangles}}

\begin{figure}[!tb]
\centering
\includegraphics[width=\columnwidth]{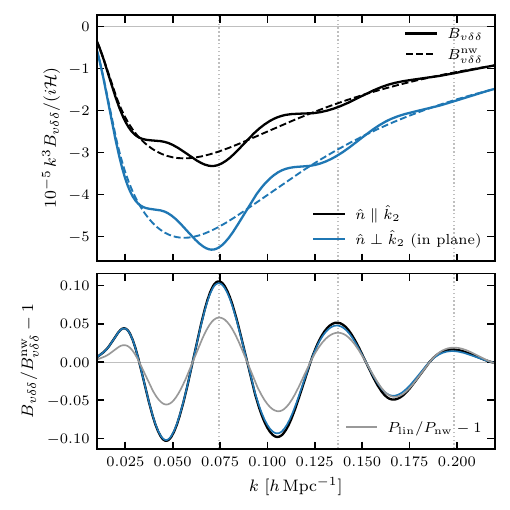}
\caption{\label{fig:vdd_fourier_equilateral}\rev{The $v\delta\delta$ bispectrum on equilateral triangles, $k_1=k_2=k_3=k$, evaluated pointwise from Eq.~\eqref{eq:eq15_closed}, for a line of sight parallel (black) and perpendicular (blue) to $\kv_2$ in the plane of the triangle. \emph{Upper panel:} $k^3B_{v\delta\delta}$ in units of $i\mathcal{H}$, with the IR-resummed linear spectrum (solid) and the no-wiggle one (dashed). \emph{Lower panel:} ratio of the two predictions minus one, compared with $P_{\rm lin}/P_{\rm nw}-1$ (grey). Dotted lines mark the peaks of $P_{\rm lin}/P_{\rm nw}$. The ratio is identical for $\theta\delta\delta$.} \textcolor{darkblue}{The perpendicular orientation gives a signal about $1.7$ times larger at the minimum. The ratio reaches $-10.3\%$ at the first trough, $k=0.047\hMpc$, against $-5.6\%$ for the power spectrum, $5.2\%$ against $3.8\%$ at $k=0.136\hMpc$, and $1.6\%$ against $1.9\%$ at $k=0.198\hMpc$. Its extrema coincide with those of $P_{\rm lin}/P_{\rm nw}$ to within $0.002\hMpc$.}}
\end{figure}

\textcolor{darkblue}{The maps of Fig.~\ref{fig:vdd_fourier_bao} retain the wiggles of $P(k_2)$ and $P(k_3)$ but average those of $P(k_1)$. To follow all three sides we evaluate the bispectrum on individual triangles with the closed form of Eq.~\eqref{eq:eq15_closed}, which carries no truncation error. We choose equilateral triangles, $k_1=k_2=k_3=k$, on which the three linear spectra oscillate in phase. The line of sight lies in the plane of the triangle and is either parallel or perpendicular to $\kv_2$.}

\textcolor{darkblue}{Figure~\ref{fig:vdd_fourier_equilateral} shows the result. The IR-resummed prediction oscillates around the no-wiggle one, with extrema at those of $P_{\rm lin}/P_{\rm nw}$. Each term of the tree-level bispectrum is a product of two linear spectra, so its relative wiggle is about twice that of a single spectrum. The IR resummation of Eq.~\eqref{eq:ir_resum} damps the oscillating part of each spectrum by a Gaussian factor in $k^2\Sigma^2$, which grows with $k$. At the first peak, $k=0.075\hMpc$, the doubling dominates, and the ratio reaches $10.5\%$ against $5.8\%$ for the linear power spectrum. By the third peak the damping has taken over, and the ratio falls below that of the power spectrum.}

\begin{figure}[!t]
\centering
\includegraphics[width=\columnwidth]{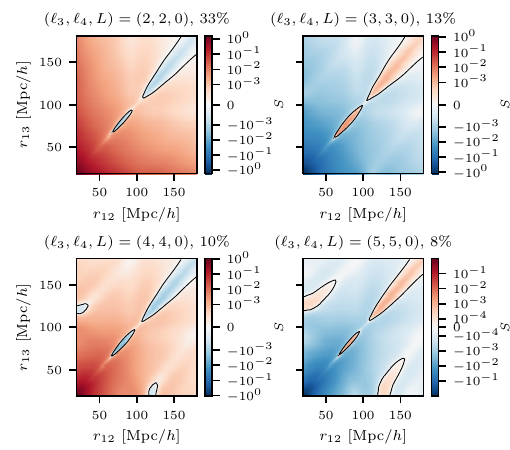}
\caption{\label{fig:tdd_config_signal}\rev{Leading coefficients \textcolor{darkblue}{$S_{\ell_3\ell_4L}(r_{12},r_{13})$} of the $\theta\delta\delta$ correlation function for the fiducial parameters, on a symmetric logarithmic colour scale covering four decades below the maximum. Black lines mark the zeros of each coefficient. The fraction of the total amplitude carried by each coefficient is given in the panel title.}}
\end{figure}

\begin{figure}[!t]
\centering
\includegraphics[width=\columnwidth]{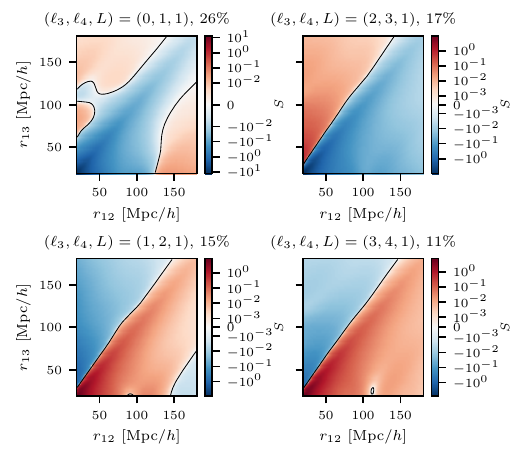}
\caption{\label{fig:vdd_config_signal}\rev{Same as Fig.~\ref{fig:tdd_config_signal} for the $v\delta\delta$ correlation function, with $\mathcal{H}=1$.}}
\end{figure}

\begin{table*}
\caption{\label{tab:ranking_config}\rev{Same as Table~\ref{tab:ranking_fourier} for the configuration-space coefficients, over $20\leq r_{12},r_{13}\leq180\Mpc$.}}
\begin{ruledtabular}
\begin{tabular}{lccccc}
$\theta\delta\delta\to v\delta\delta$ & signal & BAO & $\partial_{\ln f}$ & $\partial_{b_2}$ & $\partial_{\bG}$ \\
\hline
$(0,0,0)\to(0,1,1)$ & 0.04/0.26 & 0.10/0.31 & 0.01/0.26 & 0.19/0.24 & 0.11/0.16 \\
$(1,1,0)\to(1,2,1)$ & 0.01/0.15 & 0.23/0.19 & $<$0.01/0.14 & 0.24/0.21 & 0.37/0.23 \\
$(2,2,0)\to(2,3,1)$ & 0.33/0.17 & 0.29/0.19 & 0.19/0.17 & 0.18/0.15 & 0.20/0.17 \\
$(3,3,0)\to(3,4,1)$ & 0.13/0.11 & 0.11/0.11 & 0.07/0.11 & 0.13/0.11 & 0.11/0.12 \\
$(4,4,0)\to(4,5,1)$ & 0.10/0.08 & 0.06/0.06 & 0.06/0.07 & 0.09/0.08 & 0.07/0.08 \\
$(0,2,2)\to(0,3,3)$ & 0.04/0.01 & 0.05/0.01 & 0.13/0.02 & $<$0.01/$<$0.01 & $<$0.01/$<$0.01 \\
$L=0\to L=1$ total & 0.84/0.95 & 0.86/0.96 & 0.45/0.89 & 0.99/$>$0.99 & 0.99/$>$0.99 \\
$L=2\to L=3$ total & 0.16/0.05 & 0.14/0.04 & 0.53/0.11 & 0.01/$<$0.01 & 0.01/$<$0.01 \\
\end{tabular}
\end{ruledtabular}
\end{table*}

\textcolor{darkblue}{The velocity factor $\mu_1/k_1$ multiplies the full and the no-wiggle predictions alike, so the ratio is the same as for $\theta\delta\delta$, and the two orientations agree to within half a per cent. The bands of Fig.~\ref{fig:vdd_fourier_bao} are the counterpart of these oscillations for the two density sides. We check in App.~\ref{app:conv:fourier} that the tabulated multipoles reproduce the pointwise bispectrum on individual triangles, except near the folded limit of nearly isosceles triangles.}

\subsection{\label{sec:pred:config}\rev{Configuration space}}

\subsubsection{\label{sec:pred:config:signal}\textcolor{darkblue}{Signal hierarchy and scale dependence}}

\textcolor{darkblue}{In configuration space we follow the same path, with the amplitude of Eq.~\eqref{eq:Pamp} integrated over $20\leq r_{12},r_{13}\leq180\Mpc$. The split between sectors carries over from Fourier space, with $84\%$ of $\theta\delta\delta$ in $L=0$ and $95\%$ of $v\delta\delta$ in $L=1$. What changes is the distribution in scale and among the density-leg multipoles.}

\textcolor{darkblue}{The signal is concentrated at small separations, with $97\%$ or more of the amplitude of each $\theta\delta\delta$ coefficient at $r_{12},r_{13}<60\Mpc$. The tree-level correlation is built from products of radial transforms of the linear spectrum, which fall steeply with separation. At small separations the third side $r_{23}$ becomes short when $\rv_{12}$ and $\rv_{13}$ are nearly parallel, and the correlation grows steeply as $r_{23}$ decreases. Resolving this growth in the opening angle requires many multipoles, so the amplitude spreads along the chain $(\ell,\ell,0)$, and the monopole carries only $4\%$, against $73\%$ in Fourier space.}

\textcolor{darkblue}{This concentration describes where the amplitude lies, not where the model is most reliable, since perturbation theory is least accurate on the smallest scales. Near the diagonal $r_{12}=r_{13}$ the multipole series converges slowly, and we read the coefficients there only through integrated amplitudes (App.~\ref{app:conv:config}).}

\textcolor{darkblue}{Figure~\ref{fig:vdd_config_signal} shows the leading $v\delta\delta$ coefficients. As in $\theta\delta\delta$ (Fig.~\ref{fig:tdd_config_signal}) the amplitude lies at small separations, and as in Fourier space most coefficients with $\ell_3\neq\ell_4$ change sign across the diagonal. The pairing of Table~\ref{tab:ranking_config} is looser than in Fourier space. The term $(0,1,1)$ carries a quarter of the $v\delta\delta$ signal, although its partner $(0,0,0)$ carries $4\%$ of $\theta\delta\delta$.}

\begin{figure*}
\centering
\includegraphics[width=\textwidth]{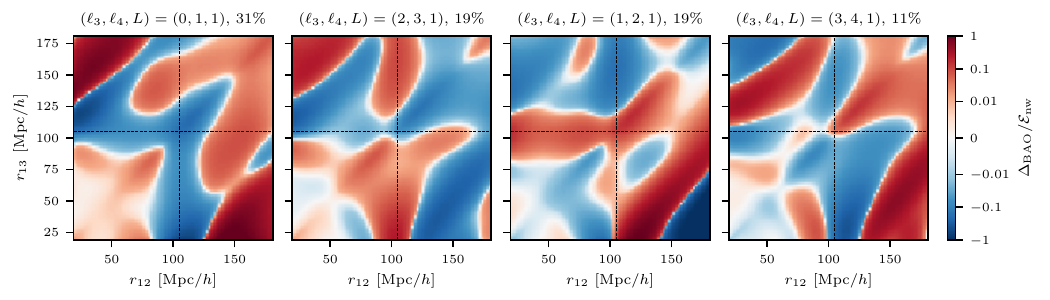}
\caption{\label{fig:vdd_config_bao}\rev{Same as Fig.~\ref{fig:vdd_fourier_bao} for the configuration-space $v\delta\delta$ coefficients of Fig.~\ref{fig:vdd_config_signal}, with the envelope smoothed over $20\Mpc$. Dashed lines mark $r_{12},r_{13}=105\Mpc$.}}
\end{figure*}

\begin{figure*}
\centering
\includegraphics[width=0.85\textwidth]{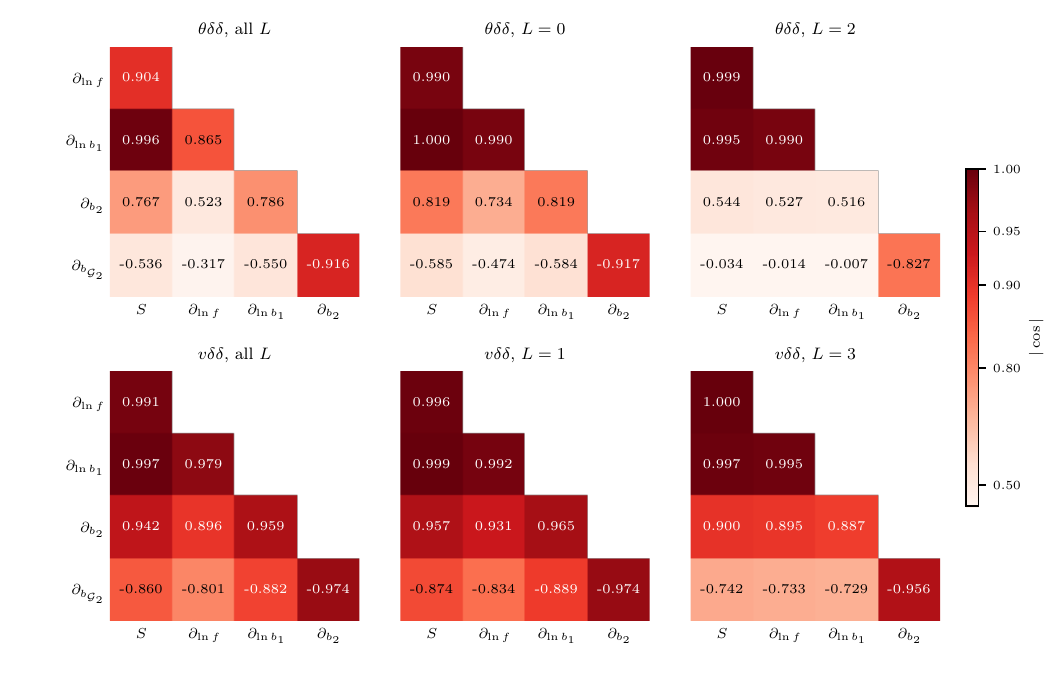}
\caption{\label{fig:similarity_config}\rev{Same as Fig.~\ref{fig:similarity_fourier} in configuration space, over $20\leq r_{12},r_{13}\leq180\Mpc$.}}
\end{figure*}

\subsubsection{\label{sec:pred:config:bao}\textcolor{darkgreen}{Baryon acoustic oscillations in the multipoles}}

\textcolor{darkblue}{Figure~\ref{fig:vdd_config_bao} shows the BAO difference with the same normalisation as Fig.~\ref{fig:vdd_fourier_bao}. In the leading coefficients it forms a cross along $r_{12}\simeq105\Mpc$ and $r_{13}\simeq105\Mpc$, where one side of the triangle crosses the acoustic scale.}

\textcolor{darkblue}{Its relative size grows with separation, because the no-wiggle correlation decays faster than the BAO feature. The difference is a few per mille of the no-wiggle coefficient where both sides are below $60\Mpc$, and reaches $4$ to $6\%$ once one side exceeds $110\Mpc$. A large local fraction is not a large share of the integrated amplitude, which remains dominated by small separations.}

\textcolor{darkblue}{\textcolor{darkgreen}{The two spaces show the feature as expected for a Fourier pair. In Fourier space it is an oscillation of a few per cent over most of the plane, whereas in configuration space it is a peak localised at the acoustic scale, large there in relative terms.} As in Fourier space, the $\theta\delta\delta$ dipole $(1,1,0)$ carries a large part of the difference (Table~\ref{tab:ranking_config}).}

\subsubsection{\label{sec:pred:config:resp}\textcolor{darkblue}{Response to the parameters}}

\textcolor{darkblue}{The parameters that set the amplitude act as in Fourier space. The derivative with respect to $\ln f$ moves amplitude to the higher sector, $\partial_{\ln f}$ and $\partial_{\ln b_1}$ are rescalings of the signal within a sector, and combining the sectors gives cosines between $\partial_{\ln f}$ and the signal of $0.90$ for $\theta\delta\delta$ and $0.99$ for $v\delta\delta$ (Fig.~\ref{fig:similarity_config}). The separation of the growth rate from the amplitude is therefore the same in both spaces.}

\textcolor{darkblue}{The second-order biases differ between the two correlators. In $\theta\delta\delta$ their responses change sign at intermediate separations, while the signal keeps a nearly constant sign, so their cosines with the signal fall below the Fourier values. In $v\delta\delta$ the two biases become more degenerate with each other than in $\theta\delta\delta$, with $|\cos|=0.97$ against $0.92$. The maps of these derivatives are produced by the figure scripts distributed with the package (App.~\ref{app:numerics}).}

\subsubsection{\label{sec:pred:triangles}\rev{Triangles of fixed shape}}

\textcolor{darkblue}{The coefficients are the natural data vector, but the BAO signal of a single triangle is easier to read in the synthesised correlation function. At tree level the correlation is a sum of terms, each depending on two sides of the triangle, so a triangle shows a BAO feature whenever one of its sides crosses the acoustic scale. \textcolor{darkgreen}{Unlike in Fourier space, where we evaluated them from the closed form, equilateral triangles cannot be built from the multipoles here, since a function of $r_{23}$ expanded in the opening angle has coefficients} that decay as $(r_{\rm min}/r_{\rm max})^{\ell}$, and on the diagonal $r_{12}=r_{13}$ they do not decay at all (App.~\ref{app:conv:config}).}

\textcolor{darkblue}{We therefore keep the shape of the triangle fixed and vary its size, with $r_{13}=3r_{12}$ and an opening angle $\theta=\pi/3$, so that $r_{23}=\sqrt{7}\,r_{12}$. For this ratio the sums at $\ell_{\max}=12$ and $14$ differ by less than $0.5\%$ of the maximum of each curve. We align $\hat{\rv}_{12}$ with the line of sight and take the velocity parallel or perpendicular to it, in the plane of the triangle, since the component normal to the plane vanishes by reflection symmetry. The parallel component is the observable $\vv\cdot\nv$. The perpendicular one is not seen by a survey and illustrates the vector correlation. In Fig.~\ref{fig:vdd_fourier_equilateral} the velocity was instead always along the line of sight, and the two curves compared two orientations of the triangle.}

\begin{figure}
\centering
\includegraphics[width=\columnwidth]{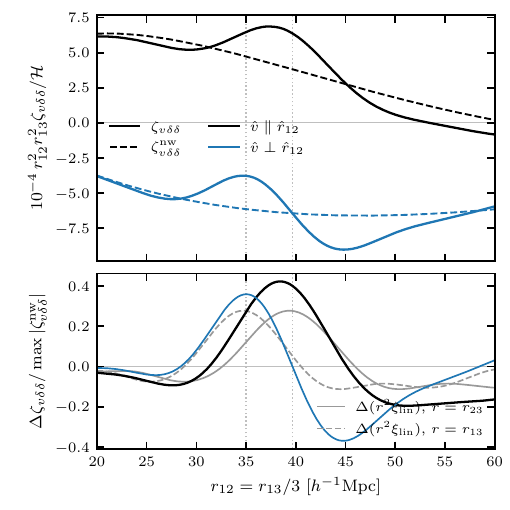}
\caption{\label{fig:vdd_config_shape}\rev{The $v\delta\delta$ correlation function on triangles of fixed shape, $r_{13}=3r_{12}$ with opening angle $\pi/3$, as a function of their size, synthesised from the multipoles with $\ell_{\max}=14$. The velocity is parallel (black) or perpendicular (blue) to $\hat{\rv}_{12}$, in the plane of the triangle. \emph{Upper panel:} $r_{12}^2r_{13}^2\zeta_{v\delta\delta}$ with $\mathcal{H}=1$, for the IR-resummed linear spectrum (solid) and the no-wiggle one (dashed). \emph{Lower panel:} difference of the two predictions, normalised to the maximum of the no-wiggle curve of each orientation. Grey lines give the same difference for the linear two-point function, $\Delta(r^2\xi_{\rm lin})/\max|r^2\xi^{\rm nw}_{\rm lin}|$, at $r=r_{23}$ (solid) and $r=r_{13}$ (dashed). Dotted lines mark the sizes at which $r_{13}$ and $r_{23}$ equal $105\Mpc$.} \textcolor{darkblue}{The two orientations give correlations of opposite sign. The parallel difference peaks at $+42\%$ for $r_{12}=38\Mpc$, where $r_{23}\simeq102\Mpc$, and reaches $-19\%$ at $r_{12}\simeq51\Mpc$. The perpendicular one peaks at $+36\%$ for $r_{12}=35\Mpc$, where $r_{13}\simeq105\Mpc$, and reaches $-37\%$ at $r_{12}=45\Mpc$.}}
\end{figure}

\textcolor{darkblue}{Figure~\ref{fig:vdd_config_shape} shows the result. The full prediction develops a bump where a side crosses the acoustic scale, followed by a dip, and the bump reaches $36$ to $42\%$ of the maximum of the no-wiggle curve. The grey curves of the lower panel show which side is responsible. The parallel bump follows the linear two-point difference at $r=r_{23}$, the perpendicular one that at $r=r_{13}$. The BAO signal of a single triangle therefore depends on which side crosses the acoustic scale and on the direction of the velocity, not on $r_{23}$ alone.}

\section{\label{sec:conclusions}Conclusions}

\textcolor{darkgreen}{The two-point velocity-density correlation measures the mean streaming motion around a single galaxy. For a pair of galaxies, the mean flow is obtained by adding the two single-galaxy contributions and a three-point term, the velocity-density-density correlation $v\delta\delta$, apart from a normalisation set by the clustering of the pair. Thus, $v\delta\delta$ supplies the additional contribution to the flow around a pair that is not specified by the two-point correlations. Since velocities respond to the gravitational pull of all matter and, on large scales, are not affected by galaxy bias, this term depends on the growth rate, galaxy bias and the amplitude of fluctuations through combinations that differ from those entering two-point statistics. In this work we have developed a model of this statistic, with the aim of comparing it with measurements from velocity and galaxy surveys.}

\textcolor{darkgreen}{We have modelled $v\delta\delta$ at the lowest order of perturbation theory at which it does not vanish, including redshift-space distortions and galaxy bias up to second order, both in Fourier and in configuration space. The model is built in two steps. We first compute the correlation of the velocity divergence with two galaxy densities, which is obtained along the same lines as the galaxy bispectrum and contains the whole dependence on cosmology and bias. The velocity statistic then follows from a purely geometric operation. Since the velocity at one vertex of the triangle depends on both of the other sides, this operation mixes the angular structure of the two density legs, but it does not depend on cosmology. The construction therefore separates a reusable geometric part, computed only once, from the cosmological dependence. We have tested the Fourier-space predictions against an exact expression and verified that the angular expansion converges in both spaces.}

\textcolor{darkgreen}{The predictions at $z=0.9$ show that, despite the many terms allowed by symmetry, the signal has a simple dependence on the line of sight. For our fiducial model, the dipole terms ($L=1$), which have the simplest line-of-sight dependence allowed for a velocity, carry $94\%$ of the squared amplitude of the signal in Fourier space and $95\%$ in configuration space. This suggests that the leading signal can be described primarily through its dipolar dependence on the line of sight, although several coefficients are still needed to resolve the dependence on the shape of the triangle, particularly in configuration space. The velocity also changes which triangles matter most. In Fourier space it favours triangles in which the velocity is carried by a long-wavelength mode, whereas in configuration space most of the amplitude lies at separations below $60\Mpc$.}

\textcolor{darkgreen}{The baryon acoustic oscillations (BAO) preserve the imprint of sound waves in the photon-baryon plasma before recombination, and they provide a further reason to study this statistic. Each term of the model contains the product of two linear power spectra, so that on large-scale equilateral triangles the relative BAO modulation is about twice as pronounced as in the power spectrum. This is a generic property of three-point statistics at this order, and the smoothing caused by large-scale bulk flows reduces the advantage on smaller scales. In configuration space the BAO appear where one side of the triangle approaches the characteristic distance travelled by these sound waves. Their signature depends both on which side crosses this scale and on the orientation of the velocity, and therefore differs from that of the galaxy three-point function, whose sides play symmetric roles.}

\textcolor{darkgreen}{The response to cosmological and bias parameters is more subtle. A change in the growth rate mostly rescales the signal, as expected since the velocity is proportional to it, but it also moves part of the signal towards the weaker terms with a more complex dependence on the line of sight. The second-order bias parameters, instead, modify the dependence on triangle shape, each in a different way. These differences make the weaker terms potentially valuable for distinguishing growth from galaxy bias despite their smaller amplitudes. They describe, however, how the predictions change, not how much can be learned from them, and the information content of $v\delta\delta$ can only be established with a covariance.}

\textcolor{darkgreen}{One of the reasons why we have devoted particular attention to configuration space is that it is the natural framework for the analysis of real catalogues. Here the survey footprint and the selection function can be accounted for by means of random catalogues, whereas in Fourier space the model would have to be convolved with the survey window, a procedure that becomes cumbersome for a three-point statistic of a vector field. For this reason, an efficient configuration-space estimator is as necessary as the model itself, and the estimators presented in this work represent a first step in this direction. We also note that our choice of modelling velocities rather than momenta does not limit the applicability of our results. At leading order, in fact, the three-point function of the density-weighted velocity field, which is the quantity effectively sampled when velocities are measured at the positions of galaxies, reduces to $v\delta\delta$ plus products of two-point correlations that are already modelled.}

\textcolor{darkgreen}{Before the model can be compared with observations, a number of issues need to be addressed. The estimators and their covariance will have to be validated against simulations, which will also clarify over which scales the lowest-order description is adequate. Observational effects specific to velocity surveys, such as measurement errors, the incomplete sampling of the velocity field and the survey geometry, will then have to be included. Once these steps are completed, the model can be applied to the growing samples of peculiar velocities obtained from distance indicators \citep{DESIRoss26,DESIDouglass26,BoruahEtAl2020,RigaultEtAl2025,CarreresEtAl2023} and to the velocity fields reconstructed from the kSZ effect \citep{SmithEtAl2018,McCarthyEtAl2025}. At higher redshifts, the dense spectroscopic samples expected from future facilities such as WST \citep{MainieriEtAl2024} will extend these analyses to much larger volumes.}

\textcolor{darkgreen}{Looking further ahead, the structure of the signal found here suggests that the model can be simplified and extended without major changes to its formulation. More generally, since $v\delta\delta$ is sensitive to how velocities respond to the density field beyond linear order, it may help to test the gravitational dynamics of structure growth and to separate it from the effects of galaxy bias. Quantifying this potential, when velocity measurements are combined with the galaxy distribution, is the natural next step, and we leave it to future work.}

\begin{acknowledgments}
\rev{We thank Michele Moresco, Alfonso Veropalumbo, Roberto Di Palma and Raul Ciancarella for useful discussions. MG and EB acknowledge support from the MIUR PRIN 2022 grant 2022NY2ZRS001 ``Optimizing the extraction of cosmological information from Large Scale Structure analysis in view of the next large spectroscopic surveys''. EB also acknowledges support from the ASI/INAF agreement n.~2024-10-HH.0 ``Scientific activity for Euclid mission, Phase E''. AN is supported by a grant from the Israel Science Foundation. This project has received funding from the European Union Horizon Europe Research and Innovation Action under grant agreement no.~101183153-WST. Views and opinions expressed are however those of the authors only and do not necessarily reflect those of the European Union or the European Research Executive Agency (REA). Neither the European Union nor the REA can be held responsible for them. Computational resources were provided by INFN Sezione di Genova, the Bologna Physics and Astronomy Department and the Leonardo supercomputing facilities at CINECA through the INAF grants INA24C3B12 and INA24C7B06. We thank the INFN IT personnel in Genova for their continuous support.}
\end{acknowledgments}

\appendix

\section{\label{app:2pt}Two-point velocity-density statistics}

\rev{For completeness, and to fix notation, we collect here the two-point cross-correlation between the peculiar velocity and the galaxy density. It enters the momentum statistics of Sect.~\ref{sec:theory:momentum} through Eq.~\eqref{eq:bisp_momentum}, and it would enter the Gaussian covariance of the three-point estimator.} Up to one loop,
\begin{equation}
    \begin{split}
    \langle v(\kv_1)\,\delta_{\rm g}(\kv_2) \rangle
    &\equiv (2\pi)^3\delta_{\rm D}(\kv_1 + \kv_2)\, P_{v\delta}(\kv_1) \, , \\
    P_{v\delta}(\kv) &= i \Hf \frac{\hat{\kv}\cdot\vv}{k}
    \left[ P^{\rm tree}_{\theta \delta}(\kv) + P^{\rm 1\text{-}loop}_{\theta \delta}(\kv) \right] \, ,
    \end{split}
    \label{eq:pvd}
\end{equation}
with the tree-level term $P^{\rm tree}_{\theta\delta}(\kv) = (b_1 + f\mu^2)\,\Plin(k)$ and the loop contributions
\begin{equation}
    \begin{aligned}
    P^{\rm 22}_{\theta \delta}(\kv) &= 2 \int_{\qv} \Plin(q)\, \Plin(|\kv - \qv|) \\
    &\quad\times Z_{\theta}^{(2)}(\qv, \kv-\qv)\, Z_{\delta}^{(2)}(\qv, \kv-\qv) \, , \\
    P^{\rm 13}_{\theta \delta}(\kv) &= 3\, Z_{\theta}^{(1)}(\kv)\, \Plin(k) \int_{\qv} Z_{\delta}^{(3)}(\kv, \qv, -\qv)\, \Plin(q) \, , \\
    P^{\rm 31}_{\theta \delta}(\kv) &= 3\, Z_{\delta}^{(1)}(\kv)\, \Plin(k) \int_{\qv} Z_{\theta}^{(3)}(\kv, \qv, -\qv)\, \Plin(q) \, ,
    \end{aligned}
    \label{eq:loop_pk}
\end{equation}
\rev{with $\int_{\qv}\equiv\int\mathrm{d}^3q/(2\pi)^3$ and symmetrised kernels.}
\rev{At tree level the real-space statistic probes the single combination $(b_1\sigma_8)(f\sigma_8)$. In redshift space the angular dependence separates it from $(f\sigma_8)^2$, which breaks the degeneracy between $b_1\sigma_8$ and $f\sigma_8$ but leaves the overall normalisation unconstrained. The loop terms add further information at the cost of the nonlinear bias parameters. A full treatment of the two-point modelling is beyond our scope, and we use the expressions above to fix conventions and to expose the dipolar mechanism that the three-point statistics of the main text generalise.}

Projecting on Legendre multipoles with respect to the line of sight, with $\vv\parallel\nv$ \rev{and $P(k,\mu)=\sum_\ell P^\ell(k)\,\mathcal{L}_\ell(\mu)$},
\begin{equation}
    P^\ell_{v\delta}(k) = \frac{i \Hf}{k} \left( P_{\theta \delta}^{\ell - 1}\, \frac{\ell}{2\ell-1} + P_{\theta \delta}^{\ell +1}\,\frac{\ell+1}{2\ell+3} \right) \, ,
    \label{eq:pvd_ell}
\end{equation}
\rev{so that only odd multipoles survive. At tree level $P_{\theta\delta}$ has a monopole $(b_1+f/3)\Plin$ and a quadrupole $(2f/3)\Plin$, and Eq.~\eqref{eq:pvd_ell} turns them into a dipole and an octupole,
\begin{equation}
    \mu\,(b_1+f\mu^2)=\Big(b_1+\frac{3}{5}f\Big)\,\mathcal{L}_1(\mu)+\frac{2}{5}f\,\mathcal{L}_3(\mu) \, .
    \label{eq:pvd_tree_ell}
\end{equation}
The dipole is the term commonly modelled in the literature \citep{AdamsBlake2017, Nusser2017}. For $b_1=2$ and $f=0.86$, the fiducial values of this work, the octupole is about $14\%$ of the dipole. This dominance is numerical rather than a formal ordering, since $f$ is of order unity, and the loop terms add higher odd multipoles.} The configuration-space multipoles follow analogously,
\begin{equation}
    \begin{split}
    \xi^\ell_{v\delta}(r) &= i\Hf \int \frac{\mathrm{d}k \, k}{2 \pi^2}\, i^{\ell} j_{\ell}(kr) \\
    &\quad\times\left( P_{\theta \delta}^{\ell -1}(k) \frac{\ell}{2\ell-1} + P_{\theta \delta}^{\ell + 1}(k) \frac{\ell + 1}{2\ell + 3} \right) \, .
    \end{split}
    \label{eq:xivd_ell}
\end{equation}

\rev{This dipolar mechanism is the two-point shadow of the $\ell_5=1$ selection rule derived in Sect.~\ref{sec:theory:tripoSH}. One velocity leg shifts every density multipole by one unit, so that the monopole of $P_{\theta\delta}$ appears as a dipole of $P_{v\delta}$ and its quadrupole feeds both the dipole and the octupole. In both cases the shift descends from Eq.~\eqref{eq:v_of_theta} alone. The mixing of adjacent orders, here $\ell\pm1$ and there the triangular chains of Eq.~\eqref{eq:bk_p_ell}, is the imprint of the vector nature of the velocity on a statistic decomposed into scalar multipoles.}

\section{\label{app:gaunt}Perturbative kernels, Gaunt integrals, and multipole derivations}

\subsection{\label{app:gaunt:kernels}Perturbative contributions, amplitudes, and kernels}

\textcolor{darkblue}{Both fields of Eq.~\eqref{eq:bias_expansion} are expanded following standard Eulerian perturbation theory in redshift space \citep{BernardeauEtAl2002},}
\begin{align}
X_{\rm g}^{(n)}(\kv)
&=
\left[
\prod_{i=1}^n
\int \frac{\mathrm{d}^3 \qv_i}{(2\pi)^3}
\,\delta^{(1)}(\qv_i)
\right]
\nonumber\\
&\quad\times
Z_{X}^{(n)}(\qv_1,\ldots,\qv_n)\,
(2\pi)^3
\delta_{\rm D}\Bigl(\kv-\sum_{j=1}^n\qv_j\Bigr),
\label{eq:pt_expansion}
\end{align}
\textcolor{darkblue}{with $X\in\{\delta,\theta\}$ and kernels $Z_\delta^{(n)}$ and $Z_\theta^{(n)}$.}

The bispectrum $\langle v(\kv_1)\delta_{\rm g}(\kv_2)\delta_{\rm g}(\kv_3)\rangle$ receives contributions from the two tree-level contractions
\begin{equation}
\begin{gathered}
\langle\theta^{(1)}(\kv_1)\,\delta^{(1)}_{\rm g}(\kv_2)\,\delta^{(2)}_{\rm g}(\kv_3)\rangle \, , \\
\langle\theta^{(2)}(\kv_1)\,\delta^{(1)}_{\rm g}(\kv_2)\,\delta^{(1)}_{\rm g}(\kv_3)\rangle \, ,
\end{gathered}
\label{eq:pt_contractions}
\end{equation}
computed following an existing kernel construction \citep{SugiyamaEtAl2021}\rev{, with both contractions of the second-order density field included, so that the result is already symmetric under $\kv_2\leftrightarrow\kv_3$}. Collecting the coefficients of each bias monomial \textcolor{darkblue}{yields the ten amplitudes $X^p$ of Eqs.~\eqref{eq:tdd_bisp} and \eqref{eq:bisp_sugylike}},
\begin{align*}
X^{(1)} &= (b_1\sigma_8)^2\sigma_8^2,
&
X^{(2)} &= (b_1\sigma_8)(f\sigma_8)\sigma_8^2,
\\
X^{(3)} &= (\bG\sigma_8^2)(b_1\sigma_8^2),
&
X^{(4)} &= (\bG\sigma_8^2)(f\sigma_8^2),
\\
X^{(5)} &= (b_2\sigma_8^2)(b_1\sigma_8^2),
&
X^{(6)} &= (b_2\sigma_8)(f\sigma_8)\sigma_8^2,
\\
X^{(7)} &= (b_1\sigma_8)^2(f\sigma_8)\sigma_8,
&
X^{(8)} &= (b_1\sigma_8)(f\sigma_8)^2\sigma_8,
\\
X^{(9)} &= (f\sigma_8)^2\sigma_8^2,
&
X^{(10)} &= (f\sigma_8)^3\sigma_8,
\end{align*}
and the associated geometric kernels, obtained by collecting the coefficients of $b_1^2$, $b_1 f$, $b_1\bG$, $\bG f$, $b_1 b_2$, $b_2 f$, $b_1^2 f$, $b_1 f^2$, $f^2$ and $f^3$ respectively,
\begin{align}
    B^{(1)} &= 2F^{(2)}_\delta(\kv_1,\kv_2) + 2G^{(2)}_\theta(\kv_2,\kv_3) \, , \label{eq:B1}\\
    B^{(2)} &= 2F^{(2)}_\delta(\kv_1,\kv_2)\,V^{(1)}_\delta(\kv_2)
          +2G^{(2)}_\theta(\kv_1,\kv_2)\,\mu_3^2 \notag\\
          &\quad+2G^{(2)}_\theta(\kv_2,\kv_3)\bigl(V^{(1)}_\delta(\kv_2)+V^{(1)}_\delta(\kv_3)\bigr) \, , \label{eq:B2}\\
    B^{(3)} &= 2K^{(2)}_\delta(\kv_1,\kv_2) \, , \label{eq:B3}\\
    B^{(4)} &= 2K^{(2)}_\delta(\kv_1,\kv_2)\,V^{(1)}_\delta(\kv_2) \, , \label{eq:B4}\\
    B^{(5)} &= 1 \, , \label{eq:B5}\\
    B^{(6)} &= V^{(1)}_\delta(\kv_2) \, , \label{eq:B6}\\
    B^{(7)} &= \frac{k_1^2\mu_2\mu_3}{k_2 k_3}
          -\frac{k_3\mu_2\mu_3}{k_2}-\frac{k_3\mu_1\mu_3}{k_1} \, , \label{eq:B7}\\
    B^{(8)} &= -V^{(1)}_\delta(\kv_1)\frac{k_3\mu_2\mu_3}{k_2}
          -2V^{(1)}_\delta(\kv_2)\frac{k_3\mu_1\mu_3}{k_1} \notag\\
          &\quad-\frac{k_3\mu_2^3\mu_3}{k_2}
          +\frac{k_1^2\mu_2\mu_3\bigl(\mu_2^2+\mu_3^2\bigr)}{k_2 k_3} \, , \label{eq:B8}\\
    B^{(9)} &= 2G^{(2)}_\theta(\kv_1,\kv_2)\,V^{(1)}_\delta(\kv_2)\,\mu_3^2 \notag\\
          &\quad+2G^{(2)}_\theta(\kv_2,\kv_3)\,V^{(1)}_\delta(\kv_2)\,V^{(1)}_\delta(\kv_3) \, , \label{eq:B9}\\
    B^{(10)} &= -V^{(1)}_\delta(\kv_1)\frac{k_3\mu_2^3\mu_3}{k_2}
          -\bigl(V^{(1)}_\delta(\kv_2)\bigr)^2\frac{k_3\mu_1\mu_3}{k_1} \notag\\
          &\quad+\frac{k_1^2\mu_2^3\mu_3^3}{k_2 k_3} \, , \label{eq:B10}
\end{align}
where $\mu_i = \hat{\kv}_i\cdot\nv$, $\mu_3$ is evaluated with $\kv_3=-(\kv_1+\kv_2)$, and the kernels read
\begin{align}
    F^{(2)}_\delta(\kv_1,\kv_2)
        &= \frac{5}{7}
          +\frac{\hat{\kv}_1\cdot\hat{\kv}_2}{2}\!\left(\frac{k_1}{k_2}+\frac{k_2}{k_1}\right)
          +\frac{2}{7}(\hat{\kv}_1\cdot\hat{\kv}_2)^2 \, , \label{eq:F2def}\\
    G^{(2)}_\theta(\kv_1,\kv_2)
        &= \frac{3}{7}
          +\frac{\hat{\kv}_1\cdot\hat{\kv}_2}{2}\!\left(\frac{k_1}{k_2}+\frac{k_2}{k_1}\right)
          +\frac{4}{7}(\hat{\kv}_1\cdot\hat{\kv}_2)^2 \, , \label{eq:G2def}\\
    V^{(1)}_\delta(\kv) &= (\hat{\kv}\cdot\nv)^2 \, , \notag\\
    K^{(2)}_\delta(\kv_1,\kv_2) &= (\hat{\kv}_1\cdot\hat{\kv}_2)^2-1 \, , \label{eq:VKdef}\\
    K^{(2)}_v(\kv_2,\kv_3) &= \mu_2 \mu_3 \, \frac{k_1^2}{2 k_2 k_3} \, . \label{eq:Kv2def}
\end{align}
We remark on the symmetric form of Eq.~\eqref{eq:Kv2def}, as its naive counterpart is a potential source of error: $K^{(2)}_v$ enters the $\theta^{(2)}\delta^{(1)}\delta^{(1)}$ contraction with an explicit factor of two counting the Wick contractions of the two linear fields inside $\theta^{(2)}$, which is only correct if the kernel is symmetric in its arguments, as is standard convention for second-order kernels. Writing it with the bare vectors, $\kv_2\cdot(\kv_2+\kv_3)$, gives a form that is \emph{not} symmetric under $\kv_2\leftrightarrow\kv_3$. Averaging the two orderings and using $\kv_2\cdot(\kv_2+\kv_3)+\kv_3\cdot(\kv_2+\kv_3)=|\kv_2+\kv_3|^2=k_1^2$ yields Eq.~\eqref{eq:Kv2def}. \rev{The kernel $K^{(2)}_v$ has a simple physical origin. It is the second-order term generated by the redshift-space mapping of the velocity divergence, $\theta_s\equiv-\nabla\cdot\mathbf{v}_s/(\Hf)$, once the Jacobian of the mapping is included. At this order the line-of-sight velocity follows from $\theta_s$ exactly as in linear theory, $\mathbf{v}_s\cdot\nv=i\Hf\,\mu_1\,\theta_s(\kv_1)/k_1$, so the velocity leg introduces no kernel beyond those of $\theta_s$.}

\rev{The tidal kernel of Eq.~\eqref{eq:VKdef} is that of the Galileon operator $\mathcal{G}_2$, so that the second-order density kernel contains $b_2/2+\bG\,[(\hat{\kv}_1\cdot\hat{\kv}_2)^2-1]$ \citep{SugiyamaEtAl2021}. A common alternative uses the traceless tidal field, whose kernel is $(\hat{\kv}_1\cdot\hat{\kv}_2)^2-1/3$. The two kernels differ by a constant, which is absorbed by $b_2$. For a term $b_K K^2$ one finds $\bG=b_K$ and $b_2=b_2^{(K)}+\tfrac43 b_K$, while for a term $\tfrac12 b_{s^2}s^2$ one finds $\bG=b_{s^2}/2$ and $b_2=b_2^{(s)}+\tfrac23 b_{s^2}$, where $b_2^{(K)}$ and $b_2^{(s)}$ are the quadratic biases in those bases. The fiducial value $\bG=-2(b_1-1)/7$ of Sect.~\ref{sec:pred} is the local Lagrangian relation in the $\mathcal{G}_2$ basis.}

\subsection{\label{app:gaunt:bpell}Derivation of the Fourier-space multipoles}

\textcolor{darkblue}{All angular integrals below reduce to Gaunt integrals,}
\begin{equation}
    \gaunt{\ell_1}{\ell_2}{\ell_3}{m_1}{m_2}{m_3} \equiv \int \mathrm{d} \Omega \, Y_{\ell_1}^{m_1}(\Omega)\, Y_{\ell_2}^{m_2}(\Omega)\, Y_{\ell_3}^{m_3}(\Omega) \, ,
    \label{eq:gaunt_def}
\end{equation}
\textcolor{darkblue}{non-vanishing only when $m_1+m_2+m_3=0$, the triangular inequality $|\ell_1-\ell_2| \le \ell_3 \le \ell_1+\ell_2$ holds, and $\ell_1+\ell_2+\ell_3$ is even. We evaluate them through the standard Wigner-$3j$ representation.}

We build the result in three steps, deferring the weight $\Klam$ until it emerges naturally. \emph{Step 1: the starting integral.} We project the closed dipolar form of Eq.~\eqref{eq:eq15_closed} onto the basis of Eq.~\eqref{eq:triposh_proj}. The dipole on the velocity leg is written through the addition theorem
\begin{equation}
\hat{\kv}\cdot\vv = \frac{4\pi}{3}\sum_{m} Y^{m}_{1}(\hat{\kv})\, Y^{*m}_{1}(\vv) \, ,
\label{eq:addition_dipole}
\end{equation}
which places one harmonic on a wavevector direction and one on $\vv$, while the scalar kernel is expanded in tables through Eq.~\eqref{eq:tdd_multipoles}. The projection then reads
\begin{widetext}
\begin{equation}
\Bt^{p \, m_3 m_4 m_5}_{\ell_3 \ell_4 \ell_5} (k_2, k_3)
= -i\Hf\,\frac{4\pi}{3}\sum_{\ell_1\ell_2,m_1m_2} B_{\ell_1\ell_2}^{p\,m_1m_2}(k_2,k_3) \!\int\! d^2\hat{\kv}_2\, d^2\hat{\kv}_3\, d^2\vv\;
\frac{k_2\,Y_1^{m}(\hat{\kv}_2)+k_3\,Y_1^{m}(\hat{\kv}_3)}{k_1^2}\,
Y_1^{*m}(\vv)\,\mathcal{Y}^{*}_{\ell_i m_i} \, ,
\label{eq:bpell_step1}
\end{equation}
\end{widetext}
where
\begin{equation*}
\mathcal{Y}^{*}_{\ell_i m_i}\equiv Y^{*m_1}_{\ell_1}(\hat{\kv}_2)\,Y^{m_2}_{\ell_2}(\hat{\kv}_3)\,Y^{*m_3}_{\ell_3}(\hat{\kv}_2)\,Y^{*m_4}_{\ell_4}(\hat{\kv}_3)\,Y^{*m_5}_{\ell_5}(\vv)
\end{equation*}
collects the table [Eq.~\eqref{eq:tdd_multipoles}] and projection harmonics and $m$ is summed over. \emph{Step 2: the factor $k_1^{-2}$.} \textcolor{darkblue}{We start from the classical expansion of $1/|\kv_2+\kv_3|$,}
\begin{equation}
     \frac{1}{k_1} = 4\pi \sum_{L=0}^{\infty} \frac{(-1)^L}{2L+1} \frac{k_<^L}{k_>^{L+1}} \sum_{M} Y_{LM}(\hat{\kv}_2)\, Y^*_{LM}(\hat{\kv}_3) \, ,
     \label{eq:one_over_k1}
\end{equation}
\textcolor{darkblue}{where $k_<=\min(k_2,k_3)$ and $k_>=\max(k_2,k_3)$, and square it. Recoupling the two harmonics on each direction with one Gaunt integral collapses them to a single pair $Y_{\Lambda M}(\hat{\kv}_2)\,Y^*_{\Lambda M}(\hat{\kv}_3)$ and defines the weight $\Klam$ of Eq.~\eqref{eq:K_Lambda},}
\begin{equation}
    \frac{1}{k_1^2} = \sum_{\Lambda=0}^{\infty} \sum_{M} \Klam(k_2,k_3)\, Y_{\Lambda M}(\hat{\kv}_2)\, Y^*_{\Lambda M}(\hat{\kv}_3) \, ,
    \label{eq:one_over_k1sq}
\end{equation}
\textcolor{darkblue}{with the internal sums restricted by $|L_1-L_2|\le\Lambda\le L_1+L_2$ and $L_1+L_2+\Lambda$ even. The quantity $\sum_{M_1+M_2=M}\bigl(\gaunt{L_1}{L_2}{\Lambda}{M_1}{M_2}{-M}\bigr)^2$ does not depend on $M$, by the completeness of the Clebsch--Gordan coefficients (we verified this numerically for several triples), so $\Klam$ depends on $\Lambda$ alone and we evaluate it at $M=0$.} \emph{Step 3: the angular integrals, one at a time.} With $\Klam$ in hand, substituting Eq.~\eqref{eq:one_over_k1sq} into Eq.~\eqref{eq:bpell_step1} turns the integrand into a product of harmonics on each of the three directions, so the solid-angle integrals decouple. We work out the first term of the numerator (the one weighted by $k_2$) and resolve its three integrals in turn. The $k_3$ term follows identically with $\hat{\kv}_2\leftrightarrow\hat{\kv}_3$. The velocity integral fixes the dipole,
\begin{equation}
\int d^2\vv\; Y_1^{*m}(\vv)\, Y^{*m_5}_{\ell_5}(\vv) = (-1)^{m_5}\,\delta_{1\ell_5}\,\delta_{m,-m_5} \, ,
\label{eq:vint}
\end{equation}
so that only $\ell_5=1$ survives and $m=-m_5$. The $\hat{\kv}_3$ integral is a product of three harmonics, that is, a single Gaunt integral [Eq.~\eqref{eq:gaunt_def}],
\begin{equation}
\begin{split}
&\int d^2\hat{\kv}_3\; Y^*_{\Lambda M}(\hat{\kv}_3)\, Y^{m_2}_{\ell_2}(\hat{\kv}_3)\, Y^{*m_4}_{\ell_4}(\hat{\kv}_3) \\
&\qquad= (-1)^{M+m_4}\,\gaunt{\Lambda}{\ell_2}{\ell_4}{-M}{m_2}{-m_4} \, ,
\end{split}
\label{eq:k3int}
\end{equation}
while the $\hat{\kv}_2$ integral carries four harmonics. Recoupling the pair $Y_{\Lambda M}\,Y_1^{-m_5}$ through an intermediate multipole $\ell'$ splits it into two Gaunt integrals,
\begin{widetext}
\begin{equation}
\int d^2\hat{\kv}_2\; Y_{\Lambda M}(\hat{\kv}_2)\, Y_1^{-m_5}(\hat{\kv}_2)\, Y^{*m_1}_{\ell_1}(\hat{\kv}_2)\, Y^{*m_3}_{\ell_3}(\hat{\kv}_2)
= \sum_{\ell'} (-1)^{M-m_5+m_1+m_3}\,\gaunt{\Lambda}{1}{\ell'}{M}{-m_5}{m_5-M}\,\gaunt{\ell'}{\ell_1}{\ell_3}{M-m_5}{-m_1}{-m_3} \, .
\label{eq:k2int}
\end{equation}
\end{widetext}
Multiplying the three results, restoring the prefactor $-i\Hf\,\tfrac{4\pi}{3}\,k_2\,\Klam(k_2,k_3)$, adding the analogous $k_3$ term, and summing over $\Lambda$, $M$ and the input pair $(\ell_1\ell_2,m_1m_2)$, reproduces the mixing matrix of Eqs.~\eqref{eq:bk_p_ell}--\eqref{eq:M_explicit}.

We stress that the second term in curly brackets is \emph{not} obtained from the first by naively relabelling $\hat{\kv}_2 \leftrightarrow \hat{\kv}_3$, $\ell_1\leftrightarrow\ell_2$, $m_1\leftrightarrow m_2$, $\ell_3\leftrightarrow\ell_4$, $m_3\leftrightarrow m_4$ in the Gaunt arguments: it must be recomputed from scratch, because the starting expression is not symmetric under $\hat{\kv}_2\leftrightarrow\hat{\kv}_3$. Two sources of asymmetry generate the sign changes $M\to-M$ and $m_2\to-m_1$ visible between the two chains of Eq.~\eqref{eq:bk_p_ell}: (i) the expansion of Eq.~\eqref{eq:one_over_k1sq} carries an \emph{unconjugated} $Y_{\Lambda M}$ on $\hat{\kv}_2$ and a \emph{conjugated} $Y^*_{\Lambda M}$ on $\hat{\kv}_3$, so that the harmonic feeding the three-harmonics integral enters with argument $-M$ in the first term (through $Y^*_{\Lambda M}=(-1)^MY_\Lambda^{-M}$) and $+M$ in the second, and (ii) the table expansion itself conjugates $Y_{\ell_1}$ but not $Y_{\ell_2}$, so that $m_2$ enters the first chain with a $+$ sign while $m_1$ enters the second with a $-$ sign. Both features are faithfully reproduced by the implementation, and their combined effect was validated against the closed form of Eq.~\eqref{eq:eq15_closed} (App.~\ref{app:conv:fourier}).

\rev{The mixing matrix of Eq.~\eqref{eq:bk_p_ell} is given explicitly by}
\begin{widetext}
\begin{align}
\label{eq:M_explicit}
\mathcal{M}^{\,m_3 m_4 m_5;\,m_1 m_2}_{\ell_3 \ell_4;\,\ell_1 \ell_2}(k_2,k_3)
&=
-i\Hf\,\frac{4\pi}{3}
\sum_{\Lambda=0}^{\infty}\sum_M
\Klam(k_2,k_3)\,
(-1)^{m_1+m_3+m_4}
\nonumber\\
&\quad\times
\Bigg\{
k_2 \sum_{\ell'}
\gaunt{\Lambda}{\ell_2}{\ell_4}{-M}{m_2}{-m_4}
\gaunt{\Lambda}{1}{\ell'}{M}{-m_5}{m_5-M}
\gaunt{\ell'}{\ell_1}{\ell_3}{M-m_5}{-m_1}{-m_3}
\nonumber\\
&\qquad
+ k_3 \sum_{\ell''}
\gaunt{\Lambda}{\ell_1}{\ell_3}{M}{-m_1}{-m_3}
\gaunt{\Lambda}{1}{\ell''}{-M}{-m_5}{M+m_5}
\gaunt{\ell''}{\ell_2}{\ell_4}{-M-m_5}{m_2}{-m_4}
\Bigg\}.
\end{align}
\begin{equation}
    \Klam(k_2,k_3) \equiv 16\pi^2 \sum_{L_1,L_2=0}^{\infty} \frac{(-1)^{L_1+L_2}}{(2L_1+1)(2L_2+1)} \frac{k_<^{L_1+L_2}}{k_>^{L_1+L_2+2}} \sum_{M_1} \left(\gaunt{L_1}{L_2}{\Lambda}{M_1}{-M_1}{0}\right)^2 \, ,
    \label{eq:K_Lambda}
\end{equation}
\end{widetext}

\textcolor{darkblue}{Here $m_5=-(m_3+m_4)$ is fixed by the velocity-leg integral. Since $\Klam$ decays as $(k_</k_>)^{\Lambda}$, the sum over $\Lambda$ converges geometrically. We set $\Lambda_{\rm max}=2\max(\ell_{\rm max}^{\rm in},\ell_{\rm max}^{\rm out})+1$, a bound fixed by the triangular inequalities of the Gaunt chains rather than a tuned cutoff, which gives $\Lambda_{\rm max}=29$ for $\ell_{\rm max}^{\rm in}=8$ and $\ell_{\rm max}^{\rm out}=14$. Each $\Klam$ is computed exactly, from the closed form in Legendre functions of the second kind near $k_2=k_3$ and from the series in $t=k_</k_>$ elsewhere. The effective truncation is therefore the output cut $\ell_{\rm max}^{\rm out}$, since high-$\Lambda$ terms feed only high $(\ell_3,\ell_4)$, and we assess convergence by varying it in App.~\ref{app:conv:fourier}.}

\subsection{\label{app:gaunt:zpell}Derivation of the configuration-space multipoles}

The configuration-space derivation follows the same three steps, starting from the direct transform of Eq.~\eqref{eq:zeta_direct} with the two plane waves Rayleigh-expanded,
\begin{equation}
{\rm e}^{i\kv\cdot\rv} = 4\pi\sum_{\ell m} i^\ell j_\ell(kr)\, Y^{*m}_{\ell}(\hat{\kv})\, Y^{m}_{\ell}(\hat{\rv}) \, ,
\label{eq:rayleigh}
\end{equation}
one per density leg. Two observations make the angular reduction identical to the Fourier one, up to the radial factors. First, the Rayleigh harmonic $Y^{*m}_{\ell}(\hat{\kv}_i)$ enters the very same solid-angle integrals over $\hat{\kv}_2$ and $\hat{\kv}_3$ as in Steps~1--3 above, so the Gaunt chains are unchanged, while its partner $Y^{m}_{\ell}(\hat{\rv}_i)$ simply carries the output multipole onto the separation directions $\hat{\rv}_{12},\hat{\rv}_{13}$. Second, and this is the only genuine difference, the weight $\Klam(k_2,k_3)$ cannot be pulled out of the wavenumber integral, since it depends on $k_<$ and $k_>$. It remains inside, alongside the Bessel functions, and it is this combination that defines the radial transform $\zeta_r(\dots;\Lambda)$ of Eq.~\eqref{eq:zeta_radial}. The three solid-angle integrals are then \emph{identical} to Eqs.~\eqref{eq:vint}--\eqref{eq:k2int}, the only new elements being the separation harmonics $Y^{m_3}_{\ell_3}(\hat{\rv}_{12})$ and $Y^{m_4}_{\ell_4}(\hat{\rv}_{13})$ that ride along from the Rayleigh expansion. Assembling them, and folding $\Klam$, the Bessel functions and the measure $k_2^2k_3^2$ into the radial transform\rev{s, together with the closure weights $k_2$ and $k_3$, which remain inside the wavenumber integral (square brackets below)}, gives the explicit form of the mixing tensor\rev{s} $\mathcal{T}\rev{^{(a)}}(\Lambda)$ of Eq.~\eqref{eq:zp_ell},
\begin{widetext}
\begin{equation}
\allowdisplaybreaks
    \begin{aligned}
    \zt^p( \vv,  \rv_{12}, \rv_{13})
    &= -\frac{i\Hf}{3\pi^3}
    \sum_{\ell_1 \ell_2 \ell_3 \ell_4} \sum_{m_1 m_2 m_3 m_4}
    \sum_{\Lambda=0}^{\infty} \sum_M (-1)^{m_1+m_3+m_4} \\
    &\quad \times \int \mathrm{d} k_2 \, \mathrm{d} k_3 \, k_2^2 k_3^2 \, \Klam(k_2,k_3)\,
    \Bigg[\,
    i^{\ell_3 + \ell_4} j_{\ell_3}(k_2 r_{12}) \, j_{\ell_4}(k_3 r_{13}) \,
    B_{\ell_1 \ell_2}^{p \, m_1 m_2}(k_2, k_3) \\
    &\quad \times Y_{\ell_3}^{m_3}(\hat{\rv}_{12}) \, Y_{\ell_4}^{m_4}(\hat{\rv}_{13}) \,
    \bigg\{
    k_2 \sum_{m_5} \sum_{\ell'}
    \gaunt{\Lambda}{\ell_2}{\ell_4}{-M}{m_2}{-m_4} \,
    \gaunt{\Lambda}{1}{\ell'}{M}{m_5}{-(M+m_5)}\,
    \gaunt{\ell'}{\ell_1}{\ell_3}{M+m_5}{-m_1}{-m_3} \,
    Y_{1}^{*m_5}(\vv) \\
    &\qquad + k_3 \sum_{m_6} \sum_{\ell''}
    \gaunt{\Lambda}{\ell_1}{\ell_3}{M}{-m_1}{-m_3}\,
    \gaunt{\Lambda}{1}{\ell''}{-M}{m_6}{M-m_6} \,
    \gaunt{\ell''}{\ell_2}{\ell_4}{-M+m_6}{m_2}{-m_4} \,
    Y_{1}^{*m_6}(\vv) \bigg\} \, \Bigg] \, ,
    \end{aligned}
    \label{eq:zpell_derivation}
\end{equation}
\end{widetext}
and relabelling the dummy velocity-leg indices with $Y_1^{*m}=(-1)^mY_1^{-m}$ so that an undaggered $Y_1^{m_5}(\vv)$ multiplies both terms, one identifies the elements of the geometric tensor\rev{s $\mathcal{T}^{(2)}$ and $\mathcal{T}^{(3)}$, multiplying the radial transforms $\zeta_r^{(2)}$ and $\zeta_r^{(3)}$ in} Eq.~\eqref{eq:zp_ell}. The relabelling phase cancels and does not survive in the final coefficient. This is a delicate point, which we verified numerically through the plane-wave identity relating $\zeta_{\ell_3\ell_4 1}$ to the direct Hankel transform of the synthesised bispectrum. As in Fourier space, the second term is derived with the roles of $\hat{\kv}_2$ and $\hat{\kv}_3$ exchanged from the outset rather than by symmetrisation, and the explicit weights $k_2$, $k_3$ are the configuration-space image of the same permutation asymmetry discussed in App.~\ref{app:gaunt:bpell}.

\textcolor{darkblue}{The tensors $\mathcal{T}^{(2)}(\Lambda)$ and $\mathcal{T}^{(3)}(\Lambda)$ are therefore the sums over $M$ of the phase $(-1)^{m_1+m_3+m_4}$ times, respectively, the first and the second Gaunt chain in curly brackets of Eq.~\eqref{eq:M_explicit}, without the prefactor and the weights $k_2$, $k_3$ and $\Klam$.}

\subsection{\label{app:gaunt:rules}Selection rules}

\rev{The mixing of Eqs.~\eqref{eq:bk_p_ell} and \eqref{eq:zp_ell} does not preserve multipole orders, since a single input pair $(\ell_1,\ell_2)$ feeds a range of output cells $(\ell_3,\ell_4)$. We collect here the conditions under which the two Gaunt chains are non-vanishing. For the first chain the azimuthal numbers satisfy $M=m_2-m_4$ and $M=m_1+m_3+m_5$, and the multipoles obey the triangular inequalities
\begin{equation}
\begin{gathered}
|\Lambda-\ell_2|\le\ell_4\le\Lambda+\ell_2\,,\qquad |\Lambda-1|\le\ell'\le\Lambda+1\,,\\
|\ell'-\ell_1|\le\ell_3\le\ell'+\ell_1\,,
\end{gathered}
\label{eq:rules_triangle}
\end{equation}
with $\Lambda+\ell_2+\ell_4$, $\Lambda+1+\ell'$ and $\ell'+\ell_1+\ell_3$ all even. The second chain follows from the exchange $\ell_1\leftrightarrow\ell_2$, $\ell_3\leftrightarrow\ell_4$ and $\ell'\to\ell''$, with $M=m_1+m_3$ and $M=m_2-m_4-m_5$. Within $\Klam$ the internal sums of Eq.~\eqref{eq:K_Lambda} run over $|L_1-L_2|\le\Lambda\le L_1+L_2$ with $L_1+L_2+\Lambda$ even.}

\rev{Two global rules follow from these conditions. The input tables are invariant under rotations about the line of sight, so that $m_1=m_2$, and the azimuthal conditions then give $m_3+m_4+m_5=0$ with $m_5\in\{-1,0,1\}$. Adding the three parity conditions of either chain gives $\ell_1+\ell_2+\ell_3+\ell_4$ odd. A Newtonian bispectrum is invariant under $\kv\to-\kv$ and has $\ell_1+\ell_2$ even, so every output cell of $v\delta\delta$ has $\ell_3+\ell_4$ odd. Relativistic corrections would enter as input tables with $\ell_1+\ell_2$ odd and would map onto $\ell_3+\ell_4$ even, which makes this rule a null test of the Newtonian model.}

\rev{The same argument fixes the multipoles of the $L$ basis of Eq.~\eqref{eq:Lbasis}. There the velocity is projected on the line of sight, $\vv=\nv$, and the line-of-sight velocity is invariant under the joint inversion of the separations and of $\nv$. The coefficients are therefore non-vanishing only for $\ell_3+\ell_4+L$ even, which for $v\delta\delta$ leaves odd $L$. For $\theta\delta\delta$ the output cells have $\ell_3+\ell_4$ even and the same argument leaves even $L$. These are the rules used in the tables of Sect.~\ref{sec:pred}.}

\section{\label{app:basis}\rev{The Slepian--Eisenstein basis}}
\begin{figure}[!t]
\centering
\includegraphics[width=\columnwidth]{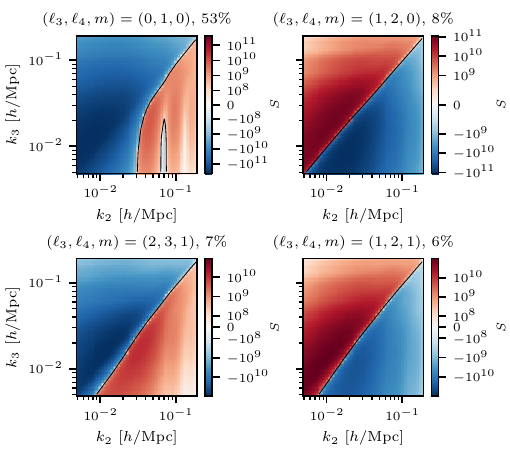}
\caption{\label{fig:vdd_fourier_se}\rev{The four leading SE coefficients of the $v\delta\delta$ bispectrum, in units of $i\mathcal{H}$, for the fiducial parameters of Sect.~\ref{sec:pred}. The percentages give the share of the total amplitude, counting both orderings of $(\ell_3,\ell_4)$ and both signs of $m$. The scale is symmetric logarithmic as in Fig.~\ref{fig:vdd_fourier_signal}, and black lines mark the zeros.}}
\end{figure}

\begin{figure}[!t]
\centering
\includegraphics[width=\columnwidth]{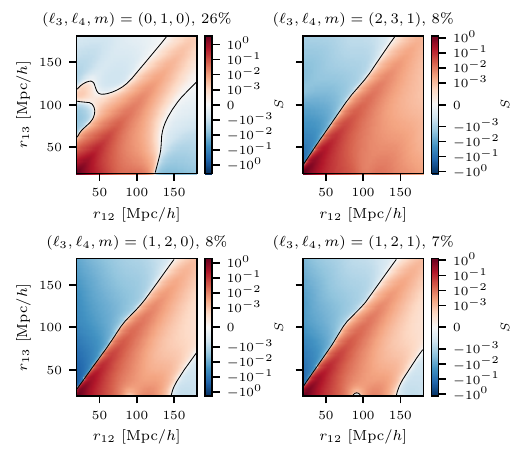}
\caption{\label{fig:vdd_config_se}\rev{Same as Fig.~\ref{fig:vdd_fourier_se} for the $v\delta\delta$ correlation function, with $\mathcal{H}=1$. The corresponding $L$-basis coefficients are shown in Fig.~\ref{fig:vdd_config_signal}.}}
\end{figure}

\rev{We derive here the coefficients $c_{\ell_3\ell_4m}$ introduced in Sect.~\ref{sec:theory:estimator} and relate them to the basis of \citep{SlepianEisenstein2018}. The configuration-space multipoles of Eq.~\eqref{eq:zp_ell} expand the correlation function on the product $Y^{m_3}_{\ell_3}(\hat{\rv}_{12})\,Y^{m_4}_{\ell_4}(\hat{\rv}_{13})\,Y^{m_5}_{1}(\vv)$. An estimator observes the velocity component along the line of sight, so we set $\vv=\nv$ and take $\nv$ as the polar axis. Since $Y^{m_5}_1(\hat{\mathbf{z}})=\sqrt{3/4\pi}\,\delta_{m_50}$ and azimuthal symmetry around $\nv$ then requires $m_4=-m_3$, the expansion reduces to}
\begin{equation}
\rev{\begin{aligned}\zeta(\rv_{12},\rv_{13};\nv)
&=\sum_{\ell_3\ell_4}\sum_{m} c_{\ell_3\ell_4m}(r_{12},r_{13})\,
\\&\quad\times Y^{m}_{\ell_3}(\hat{\rv}_{12})\,Y^{m\,*}_{\ell_4}(\hat{\rv}_{13})\,,\end{aligned}}
\label{eq:se_expansion}
\end{equation}
\rev{with}
\begin{equation}
\rev{c_{\ell_3\ell_4m}\equiv(-1)^m\sqrt{\frac{3}{4\pi}}\;\zeta^{\,m,-m,0}_{\ell_3\ell_4 1}\,,}
\label{eq:se_coeff}
\end{equation}
\rev{where the sign comes from $Y^{-m}_{\ell}=(-1)^mY^{m\,*}_{\ell}$ and $m$ runs from $-\min(\ell_3,\ell_4)$ to $\min(\ell_3,\ell_4)$. This is the expansion in Eq.~(5) of~\citep{SlepianEisenstein2018}, without its normalisation $4\pi/\sqrt{(2\ell_3+1)(2\ell_4+1)}$. \textcolor{darkblue}{The same construction applied to Eq.~\eqref{eq:triposh_proj} gives the Fourier-space coefficients, and Eq.~\eqref{eq:S_model} collects both cases.}}

\rev{Three properties follow directly from Eq.~\eqref{eq:se_expansion}. A real correlation function requires $c_{\ell_3\ell_4,-m}=c^{*}_{\ell_3\ell_4m}$, so each pair $(\ell_3,\ell_4)$ carries $\min(\ell_3,\ell_4)+1$ real parts and $\min(\ell_3,\ell_4)$ imaginary parts. A reflection through a plane containing $\nv$ maps $Y^m_\ell$ onto $Y^{m\,*}_\ell$, so the real parts are even and the imaginary parts odd under this transformation. The exchange of the two density legs, with $(\ell_3,r_{12})\leftrightarrow(\ell_4,r_{13})$, maps each coefficient onto its complex conjugate. The model is invariant under the reflection, and the imaginary parts of the computed coefficients vanish to numerical precision, so we show only the real parts.}

\rev{The $L$ basis of Eq.~\eqref{eq:Lbasis} is an orthogonal rotation of these coefficients at fixed $(\ell_3,\ell_4)$, with inverse}
\begin{equation}
\rev{c_{\ell_3\ell_4m}
=(-1)^m\sum_{L}\sqrt{\frac{2L+1}{4\pi}}
\begin{pmatrix} \ell_3 & \ell_4 & L \\ m & -m & 0 \end{pmatrix}
S_{\ell_3\ell_4L}\,,}
\label{eq:se_from_L}
\end{equation}
\rev{as follows from the orthogonality of the $3j$ symbols. It conserves the power of each pair, $\sum_L |S_{\ell_3\ell_4L}|^2=4\pi\sum_m|c_{\ell_3\ell_4m}|^2$. Under $m\to-m$ the $3j$ symbol acquires the sign $(-1)^{\ell_3+\ell_4+L}$, so the sum in Eq.~\eqref{eq:Lbasis} selects the real parts of $c_{\ell_3\ell_4m}$ for $\ell_3+\ell_4+L$ even and the imaginary parts otherwise. For $v\delta\delta$, where $\ell_3+\ell_4$ is odd, the odd $L$ of App.~\ref{app:gaunt:rules} are therefore the reflection-even content and the even $L$ the reflection-odd content, with the same counting as above. The two bases thus carry the same information, and the choice between them is a matter of estimator and of compression.}

\rev{Figure~\ref{fig:vdd_fourier_se} shows the four leading SE coefficients of $v\delta\delta$ in Fourier space, ranked with the amplitude of Eq.~\eqref{eq:Pamp} over the same ranges as in the main text. The pair $(0,1)$ has a single coefficient, equal to $-S_{011}/\sqrt{4\pi}$, and it carries the same $53\%$ of the Fourier signal and $26\%$ of the configuration-space signal. For the higher pairs the two bases differ. In the $L$ basis the term $L=1$ collects most of the power of each pair, while the SE basis spreads it over all values of $m$. For $(1,2)$ in Fourier space, $L=1$ and $L=3$ carry $13.1$ and $1.6\%$, while $m=0$ and $m=\pm1$ carry $8.3$ and $6.4\%$, with the same total $14.7\%$. As a result the four leading coefficients cover $74\%$ of the Fourier signal and $49\%$ of the configuration-space signal (Fig.~\ref{fig:vdd_config_se}) in the SE basis, against $86$ and $69\%$ in the $L$ basis. We adopt the $L$ basis in the main text for this compression, and because it also admits estimators based on fast Fourier transforms \citep{SugiyamaEtAl2019}.}

\section{\label{app:numerics}Numerical implementation}

\textcolor{newgreen}{The numerical results presented in this work have been obtained by extending to the velocity field the code \texttt{Mod3l}, presented by~\citep{FarinaEtAl2024} to model the anisotropic three-point correlation function of dark matter haloes.} The Fourier-space multipole tables $B^{p\,m_1m_2}_{\ell_1\ell_2}(k_2,k_3)$ are computed on a logarithmic grid of $N_k=256$ points over $10^{-8} \le k \le 50\hMpc$ with $N_\mu = 51$ Gauss--Legendre nodes, and input multipoles up to $\ell_{1,2}=8$. The angular projections are performed on a hybrid grid, with HEALPix pixels sampling $\hat{\kv}_3$ and Gauss--Legendre nodes sampling $\hat{\kv}_2$, and the triangle closed on $\kv_1$. \rev{Among the possible assignments of the discretisation to the three directions, this choice minimises the residual $k_2\leftrightarrow k_3$ asymmetry of the tables. It reduces the mismatch between the $B_{2,0}$ and $B_{0,2}$ multipoles from $\sim90\%$ to $\sim34\%$ at the coarsest test resolution, and the artefact decreases as the angular sampling increases.}

\textcolor{darkblue}{For the infrared resummation of Eq.~\eqref{eq:ir_resum} we use the anisotropic damping \citep{IvanovSibiryakov2018}}
\begin{equation}
  \Sigma^2(\mu) = \bigl[1+f\mu^2(2+f)\bigr]\Sigma^2_0 + f^2\mu^2(\mu^2-1)\,\delta\Sigma^2 \, ,
  \label{eq:sigma2_mu}
\end{equation}
\rev{where $\Sigma^2_0$ and $\delta\Sigma^2$ are integrals of the linear spectrum weighted by $1-j_0(q\ell_{\rm BAO})+2j_2(q\ell_{\rm BAO})$ and by $j_2(q\ell_{\rm BAO})$, with $\ell_{\rm BAO}=110\,h^{-1}{\rm Mpc}$ and $q<0.2\,h\,{\rm Mpc}^{-1}$. The growth rate in Eq.~\eqref{eq:sigma2_mu} is held at its fiducial value, as in template fits at fixed redshift. In products of two wiggly spectra the damping factors of the three legs, ${\rm e}^{-k_i^2\Sigma^2(\mu_i)/2}$, multiply each other.}

\textcolor{darkblue}{The BAO differences of Figs.~\ref{fig:vdd_fourier_bao} and \ref{fig:vdd_config_bao} are divided by the envelope of the no-wiggle coefficient $S_{\rm nw}$,}
\begin{equation}
\label{eq:envelope}
\rev{\mathcal{E}_{\rm nw}(k_2,k_3)=\big[\,G_{0.25}\ast S_{\rm nw}^2\,\big]^{1/2}(k_2,k_3)\,,}
\end{equation}
\rev{where $G_{0.25}$ is a Gaussian of width $0.25$ in $\ln k_2$ and in $\ln k_3$ and the asterisk denotes the convolution.} \textcolor{darkblue}{In configuration space the Gaussian has a width of $20\Mpc$ in $r_{12}$ and $r_{13}$.} \rev{We do not use the ratio of the full to the no-wiggle coefficient, because it diverges where the no-wiggle coefficient crosses zero. The envelope of Eq.~\eqref{eq:envelope} is positive everywhere, so the normalised difference stays finite.}

\rev{This $k$ grid is adequate in Fourier space but too coarse for the configuration-space transforms. Its spacing under-resolves the oscillations of the Bessel kernels at large separations, so that the multipoles on the diagonal $r_{12}=r_{13}$ are underestimated by an error that grows with $r$ and reaches $6\%$ of their maximum above $140\Mpc$. We therefore recompute the tables on a grid four times finer over $0.02\le k\le1\hMpc$, restricted to $10^{-4}\le k\le10\hMpc$, for a total of $516$ nodes. The restriction changes the transforms by $2\times10^{-4}$ of their maximum, and a grid eight times finer changes them by less than $0.4\%$ at all separations. The configuration-space multipoles of Sect.~\ref{sec:pred} are computed from these tables up to $\ell_{\rm max}^{\rm out}=14$, on a grid of $80$ points over $20\le r_{12},r_{13}\le180\Mpc$.}

The double Hankel transforms of Eq.~\eqref{eq:zeta_radial} are evaluated with a two-dimensional FFTLog algorithm \citep{Hamilton2000, Fang2020}, with power-law bias $\nu=1.01$ on both axes, a smoothing window applied to the highest-frequency $25\%$ of the Fourier coefficients, zero padding of 100 points per edge, and 10-point power-law extrapolation of the integrand at both ends of the $k$ range. \rev{Doubling both the padding and the extrapolation depth changes the results well below the accuracy targets of this paper and increases the cost by a factor $\sim2.5$, so we adopt the former as production settings.}

We validated the engine along two independent lines. At low multipoles ($\ell\le4$), where a direct two-dimensional quadrature of Eq.~\eqref{eq:zeta_radial} on the table grid is still reliable, the two methods agree at the sub-percent level over the full separation grid. \rev{At all multipoles we verified the stability of the FFTLog results against the ultraviolet content of the integrand. Truncating the tables above $k=1\hMpc$, where they carry less than $0.5\%$ of their diagonal power, changes the $L^2$ norm of the transformed multipoles by $1$--$4\%$, growing mildly with $\ell$ as expected from the shrinking radial window. The Fourier-space multipoles are compared with the closed form of Eq.~\eqref{eq:eq15_closed} in App.~\ref{app:conv:fourier}.}
\rev{The same truncation test exposes the failure mode of direct quadrature at high multipole order. On the table grid the local spacing at $k\simeq1\hMpc$ exceeds the oscillation period $\pi/r$ of the Bessel kernels by a factor of several for $r\gtrsim100\Mpc$. The resulting aliasing noise dominates the quadrature results, inflating the $L^2$ norms by factors of $3$--$7$ at $\ell\ge4$ and responding to the physically irrelevant ultraviolet cut by comparable factors. All configuration-space results in this paper therefore use the FFTLog engine, and we caution against quadrature-based cross-checks of double Hankel transforms at $\ell>4$ on logarithmic grids of this resolution.}

\rev{The Gaunt-chain tensors of Eqs.~\eqref{eq:bk_p_ell} and \eqref{eq:zp_ell} are purely geometric and are precomputed once per $(\ell^{\rm in}_{\rm max}, \ell^{\rm out}_{\rm max})$ pair, shared between the Fourier- and configuration-space pipelines. The configuration-space multipoles are computed once for each of the ten channels $p$ and for the two linear spectra, with and without wiggles. On the refined tables each channel took between $11$ and $27$ hours on a 32-core node, dominated by the FFTLog transforms. Any set of bias and growth parameters is then a linear combination of the stored channels through the amplitudes $X^p$, at negligible cost.}

\rev{The model is implemented in the Python package \texttt{Modv3l}\textcolor{newgreen}{, the fork of \texttt{Mod3l} that we developed for this work}, and all figures in this paper are generated by versioned scripts distributed with the package.} \textcolor{newgreen}{We plan to make the package publicly available upon publication of this paper.}

\section{\label{app:convergence}\rev{Numerical convergence and amplitude budget}}

\rev{This appendix collects the internal diagnostics of the multipole expansion. We describe the amplitude hierarchy of the multipoles, the convergence of the truncated expansion in Fourier space against the closed form, its convergence in configuration space, and the amplitude budget of a truncated data vector.}

\subsection{\label{app:conv:metric}\rev{Amplitude metric}}

\rev{We characterise the multipole content through a Parseval decomposition of the signal amplitude.} For a resummed multipole $B^{m}_{\ell_1\ell_2}(k_2,k_3)$ evaluated over the physical block $\mathcal{G}$ of the $(k_2,k_3)$ grid, its $L^2$ norm is
\begin{equation}
\label{eq:l2norm}
\big\| B^{m}_{\ell_1\ell_2} \big\|
\equiv \Bigg[\, \sum_{(k_2,k_3)\in\mathcal{G}} \big| B^{m}_{\ell_1\ell_2}(k_2,k_3) \big|^2 \,\Bigg]^{1/2}\,.
\end{equation}
\rev{Since the triple-product basis is orthonormal (Sect.~\ref{sec:theory:tripoSH}), the squared norms obey the discrete Parseval identity}
\begin{equation}
\label{eq:parseval}
\sum_{\ell_1\ell_2 m} \big\| B^{m}_{\ell_1\ell_2} \big\|^2
= \sum_{(k_2,k_3)\in\mathcal{G}} \big\langle |B|^2 \big\rangle_{\hat{\kv}}\,,
\end{equation}
\textcolor{darkblue}{so the squared norm of each multipole is its exact share of the angle-averaged power. This is an amplitude diagnostic, useful to rank multipoles and to design data vectors, and not an information metric, since only the covariance of the estimator, which is beyond our scope, can tell how much information each coefficient carries.}

\begin{figure*}
\includegraphics[width=\textwidth]{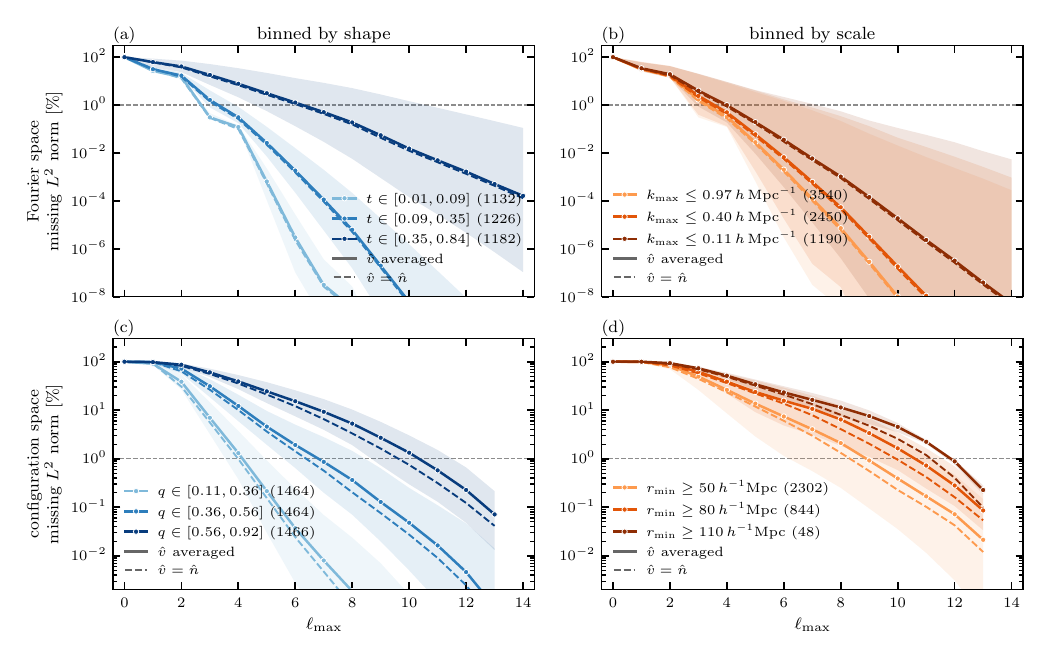}
\caption{\rev{Convergence of the truncated $v\delta\delta$ expansion, as the fraction of the $L^2$ norm still missing at a given $\ell_{\rm max}$. \emph{Top row:} Fourier space, against the direct total from the closed form of Eq.~\eqref{eq:eq15_closed}, over all triangles with $t<0.9$, binned (a) in tertiles of $t=k_</k_>$ and (b) in nested ranges of $k_{\rm max}=\max(k_2,k_3)$. Nearly equilateral triangles are excluded, since the direct total diverges there. \emph{Bottom row:} configuration space, against the sum up to $\ell^{\rm ref}_{\rm max}=14$ on the points where that sum has converged (see text), binned (c) in tertiles of $q=r_{\rm min}/r_{\rm max}$ and (d) in nested ranges of $r_{\rm min}=\min(r_{12},r_{13})$, each curve of (d) retaining triangles with both sides above $r_{\rm min}$. Solid curves average over the direction of the velocity, and dashed curves, shown as medians only, fix it along the line of sight, $\vv=\nv$. Darker colours mark the bins that converge more slowly, and the horizontal grey line marks $1\%$. The number of points is given in brackets, and bands are the $16$ to $84\%$ scatter of the solid curves within each bin.}}
\label{fig:convergence}
\end{figure*}

\subsection{\label{app:conv:fourier}\rev{Fourier space}}

\textcolor{darkblue}{Over the full physical $(k_2,k_3)$ grid the norms of Eq.~\eqref{eq:l2norm} peak in the low-order cells and decay along both axes. On the diagonal $k_2=k_3$ the mean norm per azimuthal component of the $(2,2)$ cell exceeds that of $(4,4)$ by a factor of about $4.6$, since the tree-level kernels populate only the lowest multipoles and the higher ones are generated by the resummation factor of Eq.~\eqref{eq:ir_resum}, whose expansion in powers of $k^2\Sigma^2(\mu)$ suppresses each additional order. The tables satisfy the exchange symmetry $B^{m}_{\ell_1\ell_2}(k_2,k_3)=B^{m}_{\ell_2\ell_1}(k_3,k_2)$ of Sect.~\ref{sec:theory:tdd} to better than $10^{-6}$ in relative terms. The velocity projection of Eq.~\eqref{eq:bk_p_ell} redistributes this content over the output cells $(\ell_3,\ell_4)$, which again peak at low order, as expected from the dipole shift $|\ell_1-1|\le\ell_3\le\ell_1+1$ of the selection rules.}

\rev{The truncation of this expansion can be tested against an exact reference. The full bispectrum has the closed dipolar form of Eq.~\eqref{eq:eq15_closed}, which can be evaluated at every configuration without any expansion. Beyond the low-order content of the kernels, the only source of angular structure is the expansion of $k_1^{-2}$ in Eq.~\eqref{eq:K_Lambda}, whose terms scale as $(k_</k_>)^{\Lambda}$. The residual of the truncated reconstruction therefore decays geometrically as $t^{\,\ell_{\rm max}}$, with $t=k_</k_>$ the shape parameter of the triangle.}

\rev{We test this in two ways. The first is a pointwise test on three triangles, with $t\simeq0.25$, $0.56$ and $0.69$, scanning the orientation of the velocity over $2\pi$. The truncated multipoles are synthesised into a function and compared with Eq.~\eqref{eq:eq15_closed}, which is never itself decomposed, and the comparison confirms both the limit of the series and the $t^{\,\ell_{\rm max}}$ law.} The second, shown in the upper row of Fig.~\ref{fig:convergence}, extends the test to all shapes through the Parseval identity of Eq.~\eqref{eq:parseval}. \rev{The dropped fraction $1-\|\mathcal{P}_{\ell_{\rm max}}\|^2/\|\widetilde B_{\rm dir}\|^2$ is normalised to the direct total $\|\widetilde B_{\rm dir}\|^2=\tfrac{4\pi}{3}\int d\Omega_2\, d\Omega_3\,|B_{\rm kernel}|^2/k_1^2$ from Eq.~\eqref{eq:eq15_closed}, where the integral of the exact dipole over $\vv$ supplies the factor $4\pi/3$. We verified that this total reproduces the Parseval sum of the multipoles to machine precision on converged triangles, with ratio $1.000$ for $t\lesssim0.8$. The end point of each curve is therefore a genuine residual against Eq.~\eqref{eq:eq15_closed} and not zero by construction.}
\rev{The comparison is restricted to $t<0.9$. At $k_2=k_3$ the antiparallel configuration $\hat{\kv}_2=-\hat{\kv}_3$ reaches $k_1=0$, where the kinematic factor makes $|\widetilde B_{\rm dir}|^2\propto k_1^{-2}$. The direct total then diverges logarithmically, and a global dropped fraction is not defined.}

\rev{Binned in tertiles of $t$ (panel a), the dropped fraction follows the $t^{\,\ell_{\rm max}}$ envelope, with squeezed triangles converging in a handful of multipoles and nearly equilateral ones most slowly. Nested ranges of $k_{\rm max}=\max(k_2,k_3)$ (panel b) also separate the curves, but only because a lower cut depletes the most squeezed triangles. The median $t$ rises from $0.19$ for the full sample to $0.25$ for $k_{\rm max}\le0.40\hMpc$ and $0.35$ for $k_{\rm max}\le0.11\hMpc$. This binning therefore restates the dependence on shape, and we read the Fourier convergence in terms of $t$.}

\rev{The dashed curves fix the velocity along the line of sight, $\vv=\nv$, so that only the rows with $m_5=0$ survive. Their direct total is $\int d\Omega_2\, d\Omega_3\,(\hat{\kv}_1\cdot\nv)^2\,|B_{\rm kernel}|^2/k_1^2$, with no average over $\vv$, and it again matches the Parseval sum to machine precision for $t<0.5$. In Fourier space the two orientations converge at the same rate. In the slowest tertile of $t$ the median missing fraction for $\vv=\nv$ is $12$ to $17\%$ lower at every $\ell_{\rm max}$, and the $1\%$ threshold is reached at the same order in every bin, so we read the Fourier convergence as set by the shape alone.}
\rev{This agrees with the structure of the mixing. The factor $k_1^{-2}$ is homogeneous of degree $-2$, and its Legendre coefficients depend on $t$ alone, $\Klam\propto t^{\Lambda}$. The number of multipoles required is then a function of shape with no dependence on the overall wavenumber. This conclusion rests on the homogeneity of $k_1^{-2}$ and on the measured shift of $t$ across the bins, not on a scan at fixed $t$, and we contrast it with configuration space in App.~\ref{app:conv:config}.}

\rev{A scan of the opening angle at fixed $(k_2,k_3)$ locates the residual of the truncation. For triangles with $t\simeq0.54$ and $0.32$, the synthesis at $\ell_{\max}=14$ agrees with Eq.~\eqref{eq:eq15_closed} within $3\%$ of the maximum over the opening angle $\theta$ between $\kv_2$ and $\kv_3$. For the nearly isosceles triangle $(k_2,k_3)=(0.09,0.11)\hMpc$, with $t\simeq0.84$, the agreement holds for $\theta<0.8\pi$, but the folded peak is still underestimated by $27$ to $47\%$. The peak has an angular width $\Delta\theta\simeq(k_3-k_2)/\sqrt{k_2k_3}\simeq0.2$, which requires multipoles up to $\ell\sim\pi/\Delta\theta\simeq16$. This is why we evaluate the equilateral triangles of Sect.~\ref{sec:pred:fourier_triangles} directly from the closed form.}

\subsection{\label{app:conv:config}\rev{Configuration space}}

\rev{Each multipole of Eq.~\eqref{eq:zp_ell} is exact, but the truncated expansion of Eq.~\eqref{eq:zeta_multipole_structure} converges to the full function through a different mechanism than in Fourier space. Each multipole now integrates over all wavenumbers, and the spherical Bessel function $j_\ell(kr)$ has support up to $\ell\simeq kr$. The number of multipoles needed at separation $r$ is therefore set by the product of $r$ with the wavenumbers that carry power, a criterion of scale with no geometric suppression. At $r\simeq100\Mpc$, with power at $k\simeq0.1$ to $0.3\hMpc$, this gives $\ell\sim10$ to $30$.}
\rev{This is a property of the triple-product basis and not of the model. In an analysis multipole by multipole every retained coefficient is predicted exactly, and the truncation only sets the share of the total amplitude that the data vector captures. We quantify this share in App.~\ref{app:conv:budget}.}

\rev{All results in this subsection use the refined tables of App.~\ref{app:numerics}, on $80$ points over $20\le r_{12},r_{13}\le180\Mpc$, with $\ell^{\rm out}_{\rm max}=14$. The $\theta\delta\delta$ multipoles follow from the one-to-one transform of Eq.~\eqref{eq:zeta_tdd}, and the metric of Eqs.~\eqref{eq:l2norm} and \eqref{eq:parseval} carries over with $(k_2,k_3)$ replaced by $(r_{12},r_{13})$ and $B$ by $\zeta$.}

\textcolor{darkblue}{The transform flattens the hierarchy. Per azimuthal component, the diagonal cells $(\ell,\ell)$ fall by a factor of only $1.7$ from $\ell=2$ to $\ell=8$, against more than $20$ in Fourier space. Indeed, the filter $j_\ell(kr)$ peaks at $kr\simeq\ell$, so higher multipoles weight the wavenumbers $k\simeq\ell/r$, which at $r=20\Mpc$ still overlap the region where the model carries power. Summed over $m$, the same cells even grow with $\ell$, while at $\sqrt{r_{12}r_{13}}>60\Mpc$ the $(2,2)$ cell leads more clearly, with the higher cells at about $0.55$ of it per component.}

\rev{The $v\delta\delta$ multipoles follow by contracting the radial transforms of Eq.~\eqref{eq:zeta_radial} with the mixing tensors of Eq.~\eqref{eq:zp_ell}. Projecting the synthesised $\zeta(\vv)$ onto harmonics of the velocity direction at fixed $(\rv_{12},\rv_{13})$ returns a pure $\ell_5=1$ signal, with the $\ell_5=0$ and $\ell_5=2$ projections zero to machine precision, as anticipated in Sect.~\ref{sec:config:structure}. The amplitude map peaks on the pair $(0,1)$, with $14\%$ of the total per component, and decays slowly along the band $\ell_4=\ell_3\pm1$, which inherits the flattening of the $\theta\delta\delta$ input.}

\rev{No pointwise closed form is available in configuration space. The direct object is the six-dimensional integral of Eq.~\eqref{eq:zeta_direct}, and a Monte Carlo evaluation of it is defeated by the oscillating phases $e^{i\kv\cdot\rv}$ and by the factor $k_1^{-2}$, which make its variance divergent. The multipoles are instead exact double Hankel transforms of the Fourier multipoles tested above, so the only residual is the truncation in $\ell$.}
\rev{By the angular Parseval identity the orientation-averaged power at fixed $(r_{12},r_{13})$ is $\sum_{\rm out}\|\zeta_{\rm out}\|^2$. We estimate the total from the multipoles up to $\ell^{\rm ref}_{\rm max}=14$, which equals the true total only where the series has converged at that order. We therefore keep only the points where the last two shells, $\ell=13$ and $14$, carry less than $1\%$ of the total. These cover $69\%$ of the grid, up to $\sqrt{r_{12}r_{13}}\simeq141\Mpc$, and on them the dropped fraction $1-\|\mathcal{P}_{\ell_{\rm max}}\|^2/\|\zeta_{\rm dir}\|^2$ is converged to within $1\%$.}

\rev{The lower row of Fig.~\ref{fig:convergence} shows the result. In nested ranges of $r_{\rm min}$ (panel d), the convergence slows as $r_{\rm min}$ grows. The median missing fraction falls below $1\%$ at $\ell_{\rm max}=9$ for $r_{\rm min}\ge50\Mpc$, at $11$ for $r_{\rm min}\ge80\Mpc$ and at $12$ for $r_{\rm min}\ge110\Mpc$. Binned instead in tertiles of the shape $q=r_{\rm min}/r_{\rm max}$ (panel c), elongated triangles converge fastest and nearly isosceles ones most slowly, the counterpart of the ordering in $t$ in Fourier space. The two variables are correlated but distinct, and both matter in configuration space.}

\rev{With the velocity along the line of sight (dashed curves) the configuration-space multipoles converge faster. At $\ell_{\rm max}=8$ the median missing fraction is $1.3$, $4.1$ and $8.0\%$ for the three cuts in $r_{\rm min}$, against $2.1$, $6.5$ and $11.4\%$ for the average over $\vv$, and the $1\%$ threshold moves from $11$ to $10$ for $r_{\rm min}\ge80\Mpc$. The average also contains the components of $\vv$ transverse to $\nv$, so the two curves describe different quantities and not two bases of one quantity. Both are evaluated on the same points, those where the sum up to $\ell^{\rm ref}_{\rm max}$ has converged for either orientation.}

\rev{The dashed curves are also what the line-of-sight bases measure. At fixed $(\ell_3,\ell_4)$ the coefficients $c_{\ell_3\ell_4m}$ of App.~\ref{app:basis} and the $L$ basis are orthogonal recombinations of the $m_5=0$ rows, and they keep the power of each pair unchanged. The curves are therefore identical in the three descriptions, as long as the truncation runs over $\max(\ell_3,\ell_4)$ and no weight is applied per pair. A truncation in $L$, or the normalisation $4\pi/\sqrt{(2\ell_3+1)(2\ell_4+1)}$ of \citep{SlepianEisenstein2018}, would change them, and we keep the unweighted form throughout.}

\rev{The diagonal $r_{12}=r_{13}$ is excluded by construction. There the last two shells retain between $1$ and $19\%$ of the power, because equal legs spread it most uniformly over the $(\ell_3,\ell_4)$ plane, as the equilateral limit does in Fourier space. The retained points therefore stop at $q=0.92$. For the same reason the triangles of fixed shape in Sect.~\ref{sec:pred:triangles} have $r_{13}=3r_{12}$, where the sums at $\ell_{\rm max}=12$ and $14$ agree within $0.5\%$ (Fig.~\ref{fig:vdd_config_shape}).} \textcolor{darkblue}{In the same region the coefficients $(\ell,\ell,0)$ of Fig.~\ref{fig:tdd_config_signal} show islands of the opposite sign, at $r\simeq50$ to $90\Mpc$ and beyond $110\Mpc$. They are present with the same shape in the no-wiggle prediction, so they do not originate from the BAO oscillations, and they do not affect the integrated amplitudes, which are dominated by small separations.}

\rev{That convergence is fastest at small separations, where the field is most nonlinear, does not mean that the physics is simpler there. The perturbative expansion lives in the kernels and in the radial amplitude, while the angular expansion truncated here is kinematic. Its order is set by $\ell\sim k_{\rm peak}r$, so a large triangle excites many harmonics and a small one few. A truncated $v\delta\delta$ analysis is thus cheapest on the small scales where the dynamics is hardest, and most demanding on the acoustic scales where the dynamics is easiest.}
\rev{The same argument explains why the correspondence $r\leftrightarrow1/k$ between the two spaces is not exact. Configuration space carries the fixed scale $k_{\rm peak}$ of the power spectrum, which makes convergence depend on the absolute separation. At fixed moduli Fourier space retains only the dimensionless shape $t$, so convergence is controlled by a scale in one space and by a shape in the other.}

\subsection{\label{app:conv:budget}\rev{Amplitude budget of the truncated expansion}}

\rev{Each retained multipole is exact, but a data vector truncated at $\ell_{\rm max}$ captures only part of the total squared norm of Eq.~\eqref{eq:parseval}. Table~\ref{tab:completeness} reports the share carried by the cells with $\max(\ell_a,\ell_b)\le L$, for grids restricted to $r_{12},r_{13}\ge r_{\rm min}$. For $\theta\delta\delta$ the transform is diagonal, so the budget reflects the multipole content of the model, normalised to the input order $\ell=8$. For $v\delta\delta$ it is normalised to $\ell^{\rm ref}_{\rm max}=14$, and since the series has not converged on the diagonal, its entries are upper bounds.}

\begin{table}
\caption{\label{tab:completeness}\rev{Share of the total $L^2$ norm of the configuration-space multipoles captured by the cells with $\max(\ell_a,\ell_b)\le L$, for grids restricted to $r_{12},r_{13}\ge r_{\rm min}$ (in $h^{-1}\,{\rm Mpc}$). The $\theta\delta\delta$ shares are normalised to $\ell=8$, the order of the input tables, and the $v\delta\delta$ shares to $\ell=14$.}}
\begin{ruledtabular}
\begin{tabular}{llcccc}
 & $r_{\rm min}$ & $L\le2$ & $L\le4$ & $L\le6$ & $L\le10$ \\
\hline
$\theta\delta\delta$ & 20 & 12.3\% & 35.2\% & 66.0\% & \\
 & 40 & \phantom{0}7.0\% & 23.2\% & 55.1\% & \\
 & 60 & \phantom{0}4.9\% & 17.4\% & 49.3\% & \\
 & 80 & \phantom{0}2.9\% & 13.2\% & 41.9\% & \\
\hline
$v\delta\delta$ & 20 & 35.6\% & 66.0\% & 82.0\% & 96.6\% \\
 & 40 & 17.2\% & 46.1\% & 67.4\% & 91.9\% \\
 & 60 & 10.9\% & 34.7\% & 57.1\% & 87.3\% \\
 & 80 & \phantom{0}8.2\% & 28.4\% & 49.6\% & 83.3\% \\
\end{tabular}
\end{ruledtabular}
\end{table}

\rev{The completeness decreases as $r_{\rm min}$ grows in both correlators, because the support $\ell\lesssim kr$ widens at larger separations. This is the opposite of Fourier space, where convergence is set by shape and does not depend on scale. For $r_{\rm min}=80\Mpc$ a $v\delta\delta$ data vector truncated at $\ell\le4$ captures $28\%$ of the amplitude and one truncated at $\ell\le10$ captures $83\%$. The two correlators are normalised to different orders, so their columns should not be compared directly.}
\rev{These shares measure amplitude left out of a truncated data vector, not a bias, since every retained coefficient is exact. The noise of high-order multipoles also grows with the normalisation of the estimator, so the omitted coefficients are the ones measured worst. The information actually lost requires the Gaussian covariance of the $v\delta\delta$ estimator, which we defer to future work (Sect.~\ref{sec:conclusions}).}

\bibliography{bibliography}

\end{document}